\documentclass[aps,prb,reprint,superscriptaddress]{revtex4-2}

\usepackage{amsmath,bm,mathrsfs,color}
\usepackage{graphicx}
\usepackage{enumerate}

\begin{document}

% Use the \preprint command to place your local institutional report
% number in the upper righthand corner of the title page in preprint mode.
% Multiple \preprint commands are allowed.
% Use the 'preprintnumbers' class option to override journal defaults
% to display numbers if necessary
%\preprint{}

%Title of paper
\title{Long-range permeation of wave function and superficial surface state due to strong quantum size effects in topological Bi/BiSb heterojunctions}

% repeat the \author .. \affiliation  etc. as needed
% \email, \thanks, \homepage, \altaffiliation all apply to the current
% author. Explanatory text should go in the []'s, actual e-mail
% address or url should go in the {}'s for \email and \homepage.
% Please use the appropriate macro foreach each type of information

% \affiliation command applies to all authors since the last
% \affiliation command. The \affiliation command should follow the
% other information
% \affiliation can be followed by \email, \homepage, \thanks as well.
\author{Yuya Asaka}
\author{Tatsuki Kikuchi}
%\email[]{Your e-mail address}
%\homepage[]{Your web page}
%\thanks{}
%\altaffiliation{}
\affiliation{Department of Engineering Science, University of Electro-Communications, Chofu, Tokyo 182-8585, Japan}
\author{Yuki Fuseya}
\affiliation{Department of Engineering Science, University of Electro-Communications, Chofu, Tokyo 182-8585, Japan}
\affiliation{Institute for Advanced Science, University of Electro-Communications, Chofu, Tokyo 182-8585, Japan}

%Collaboration name if desired (requires use of superscriptaddress
%option in \documentclass). \noaffiliation is required (may also be
%used with the \author command).
%\collaboration can be followed by \email, \homepage, \thanks as well.
%\collaboration{}
%\noaffiliation

\date{\today}

%%%%%%%%%%%%%%%%%%%%%%%%%%%%%%%%%%%%%%%%%%%%%%%%%%%%%%%%%%%%%%%%%%%%%%%%
\begin{abstract}
% insert abstract here

The quantum size effect has a significant impact on electrons, such that it can even change their topologically protected properties. An example of this phenomenon is the gap opening in the topologically protected gapless surface state in finite-thickness topological-insulator films. However, much is not known about the quantum size effect in topological heterojunctions.
In this study, by calculating the single-particle spectrum of Bi/Bi$_{1-x}$Sb$_x$ based on the well-known Liu-Allen model, we found that the strong quantum size effect in topological heterojunctions yields an unexpected band alignment. The wave functions permeate each other through the attached materials, and this occurs even in 80-nm-thick heterojunctions.
Furthermore, we theoretically found that one of the two major spectra obtained from the Bi surface does not represent the true surface state of Bi. This finding may overturn the previous understanding of the topological surface state of Bi.
%We believe that the long-range permeation observed in this study could lay the foundations for the fabrication of novel ``wave-functional" alloys without requiring the introduction of randomly distributed atoms. 

\end{abstract}
%%%%%%%%%%%%%%%%%%%%%%%%%%%%%%%%%%%%%%%%%%%%%%%%%%%%%%%%%%%%%%%%%%%%%%%%

% insert suggested keywords - APS authors don't need to do this
%\keywords{}

%\maketitle must follow title, authors, abstract, and keywords
\maketitle

% body of paper here - Use proper section commands
% References should be done using the \cite, \ref, and \label commands
%%%%%%%%%%%%%%%%%%%%%%%%%%%%%%%%%%%%%%%%%%%%%%%%%%%%%%%%%%%%%%%%%%%%%%%%

\section{Introduction}
\begin{figure}[h]
 \includegraphics[width=8cm]{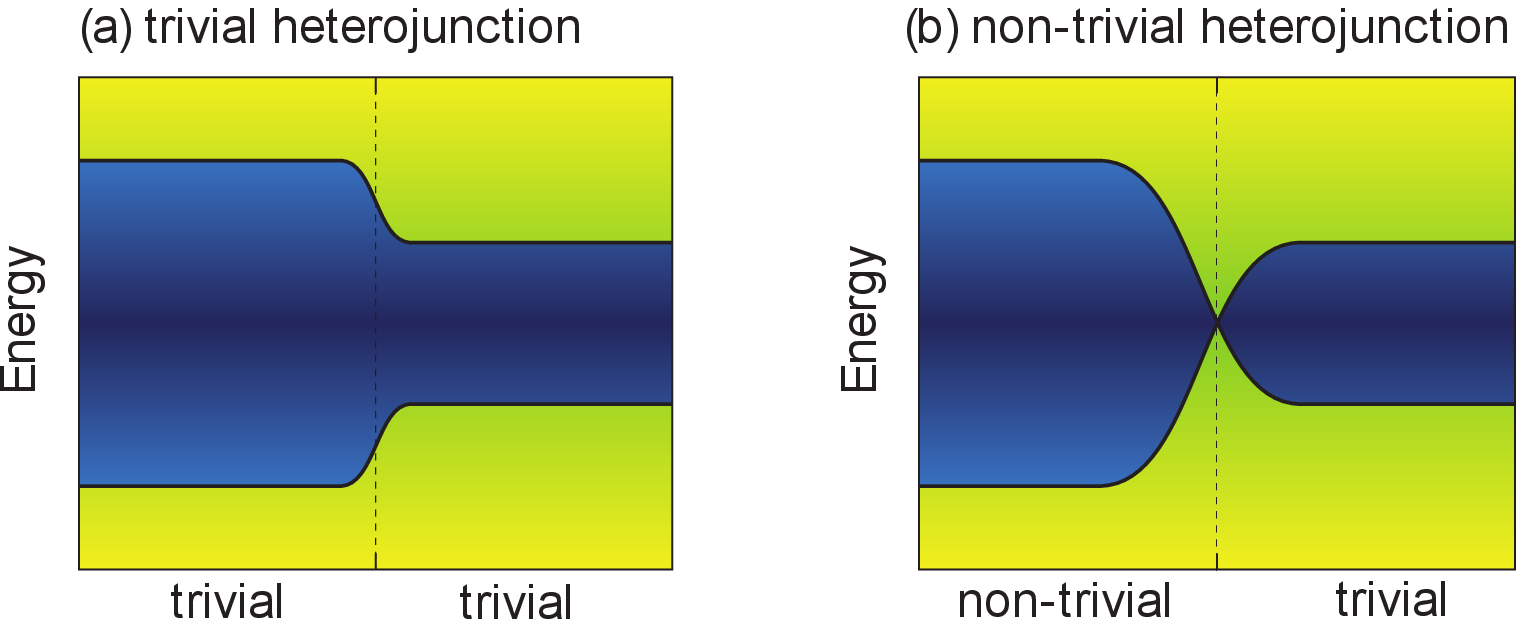}
 \caption{\label{Fig0} Schematics of heterojunctions (a) between topologically trivial semiconductors and (b) between trivial and nontrivial semiconductors. 
In finite-thickness heterojunctions, these band alignments are discretized due to the quantum size effect.
}
\vspace{5mm}
\includegraphics[width=8cm]{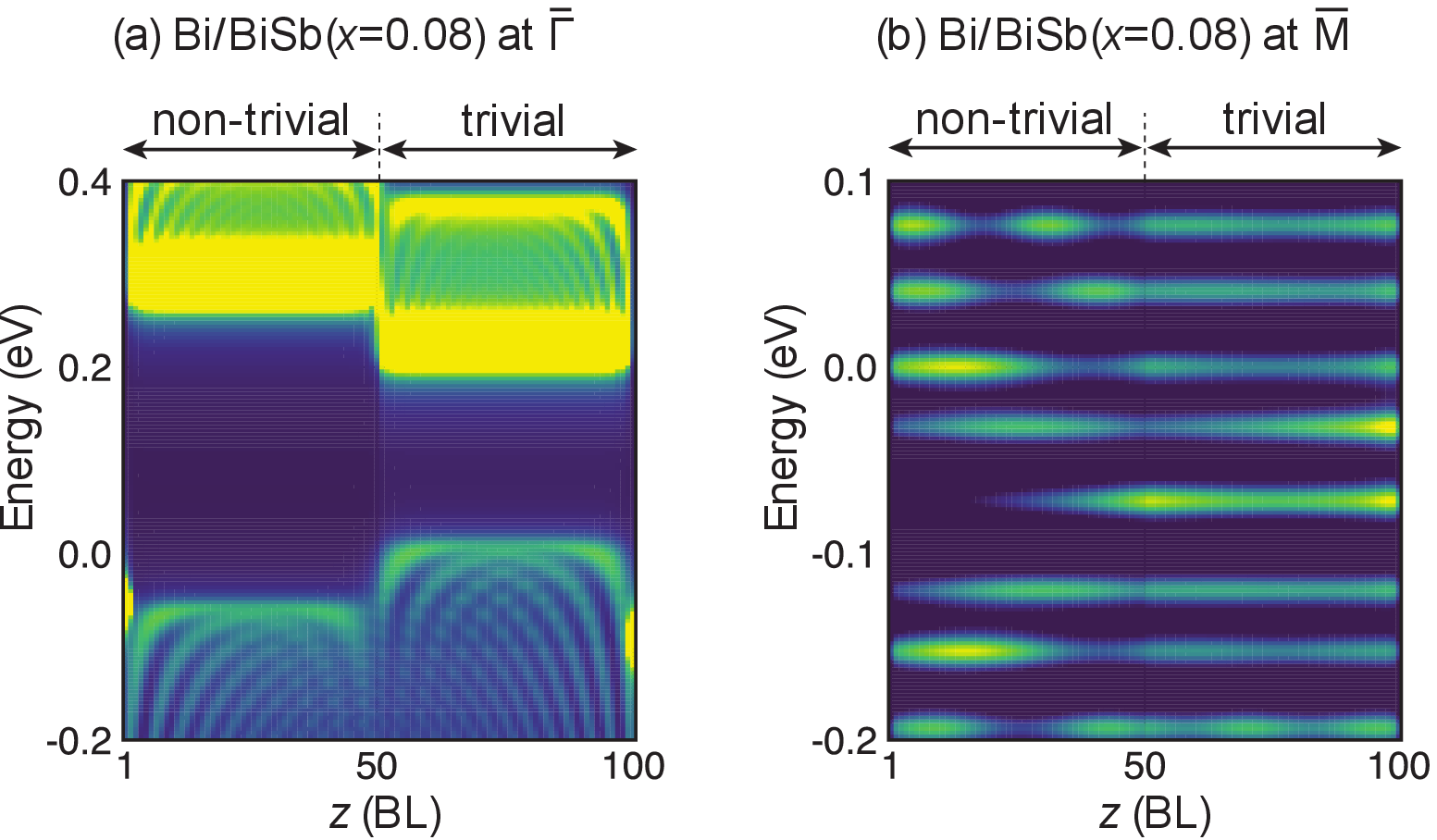}
 \caption{\label{Fig02} Present results of topological heterojunction Bi/Bi$_{1-x}$Sb$_x$ ($x=0.08$) of 50+50 BL at (a) $\bar{\Gamma}$-point and (b) $\bar{M}$-point.}
 \end{figure}

The design and manipulation of heterojunctions, which are interfaces between two dissimilar semiconductors, is one of the most effective ways of controlling band structure. Solar cells, lasers, and high-electron-mobility transistors are examples of successful applications of band engineering via heterojunctions. In general, the band alignment of a heterojunction can be described as a natural connection between the band edges of two semiconductors, as illustrated in Fig. \ref{Fig0} (a) \cite{Kittel_book_intro,Ihn_book,Bastard_book}.
In the mid-1980s, the possibility of an interface state (IS) in the gap in a heterojunction between mutually band-inverted semiconductors, such as Pb$_{1-x}$Sn$_x$Te \cite{Volkov1985,Fradkin1986} and Hg$_{1-x}$Cd$_x$Te \cite{Cade1985,Chang1985,LinLiu1985,Pankratov1987}, was proposed.
In recent years, the appearance of the IS has been reexamined in terms of topology.
When a topologically nontrivial insulator faces a topologically trivial insulator (including a vacuum), a gapless conducting state must exist at the interface in order to connect bands with the same symmetry, as shown in Fig. \ref{Fig0} (b)\cite{Hasan2010,XLQi2011}. 
The heterojunction of Pb$_{1-x}$Sn$_x$Te and Hg$_{1-x}$Cd$_x$Te can be regarded as a topological heterojunction from a contemporary viewpoint \cite{Bernevig2006,Hsieh2012,GWu2013,Shoman2015,Krizman2022}. 

%======================================================================== 
%========================================================================

The above arguments regarding the topological IS assume an infinite-thickness heterojunction, i.e., a junction between semi-infinite-thickness semiconductors. 
However, in reality, heterojunctions are fabricated by stacking finite-thickness films.
It is naively expected that the band alignment shown in Fig. \ref{Fig0} is discretized owing to the quantum size effect in the heterojunctions of finite-thickness.
In addition,
the quantum size effect changes the topological surface state (SS) of finite-thickness topological insulators \cite{BZhou2008,Linder2009,HZLu2010,SQShen_book,Fuseya2018}; the gapless metallic SS in a semi-infinite system produces a gap in the finite-thickness system owing to the interference between the surfaces of the opposite sides.
The quantum size effect becomes substantially prominent when the wavelength of the electrons is increased. Even the topologically protected conditions can be changed by the quantum size effect arising from long-wavelength electrons, such as in Bi$_{1-x}$Sb$_x$ (BiSb) \cite{Fuseya2018,Aguilera2021}.
Similar quantum size effects are expected in topological heterojunctions; the gapless IS of topologically non-trivial heterojunction, Fig. \ref{Fig0} (b), may be changed into an ordinary IS of trivial heterojunction, Fig. \ref{Fig0} (a).
%\textcolor{red}{
%For example, in the superlattice, composed of thin films of a magnetically doped topological insulator and ordinary insulator, it was theoretically shown that the proximity effect of wavefunctions can change the topology of the system \cite{Burkov2011}.
%}
However, it is not clearly understood how the quantum size effect changes the band alignment in topological heterojunctions.
Several works have reported the topological ``proximity" effect. The topological SS seeped out the attached topologically trivial atomic layer in Zn{\it M}/Bi$_2$Se$_3$ ({\it M}=S, Se, Te)  \cite{GWu2013} and in Bi/TlBiSe$_2$\cite{Shoman2015}. Meanwhile, the interfacial state seeped out the attached topological insulator in Bi$_2$Se$_3$/MnSe \cite{Eremeev2013} and in MnBi$_2$Se$_4$/Bi$_2$Se$_3$ \cite{Hirahara2017}.
In these works, however, the penetration of wavefunction is limited only in close proximity of $\sim 1$ nm.
%In these works, the proximity effect the attached trivial material is only one-atomic-layer thick. 
It is difficult to investigate how the quantum size effect affects the band alignment of topological heterojunctions in such extremely narrow region. %, because the surface, interface, and bulk states are not well separated.
%Recently, the topological heterostructure of Pb$_{1-x}$Sn$_x$Se/Pb$_{1-y}$Eu$_y$Se was investigated and interactions between 

Another type of topological proximity effect can be also observed in the topological superlattice \cite{Burkov2011}. It is composed of thin films of a magnetically doped topological insulator and ordinary insulator. The proximity effect of wavefunctions can change the topology of the system.

In this study, we investigate the ``thick" topological heterojunction between Bi and BiSb, where the surface, interface, and bulk states are spatially well separated. We demonstrate that the band alignment of the topological heterojunction can be neither Fig. \ref{Fig0} (a) nor (b), contrary to theoretical expectations. The strong quantum size effect yields an unexpected band alignment [Fig. \ref{Fig02} (b)] that cannot be classified into the typical category of heterojunctions \cite{Kittel_book_intro,Ihn_book,Bastard_book}. 

The Bi/BiSb heterojunction provides an ideal foundation for investigating the quantum size effect in heterojunctions. 
Bi is well known for its exceptionally long wavelength ($\sim 100$ nm) \cite{Liu1995,ZZhu2011,Fuseya2015} that induces a strong quantum size effect \cite{Fuseya2018,Aguilera2021}.
Another advantage of this system is that we can control the topology of one side of the junction, BiSb, by slightly changing the Sb content, thereby keeping the crystal structure and the lattice constant almost unchanged. It is experimentally well established that the conduction band (CB) and valence band (VB) at the $L$-point in the bulk Brillouin zone are inverted at $x=x_c^{\rm exp} \simeq 0.04$ \cite{Lerner1968,Tichovolsky1969,Brandt1972,Oelgart1976,Lenoir1996} accompanied by the $Z_2$-topological transition \cite{Fu2007,Hasan2010}.

%%%%%%%%%%%%%%%%%%%%%%%%%%%%%%%%%%%%%%%%%%%%%%%%%%%%%%%%%%%%%%%%%%%%%%%%
\section{Theory}
In this study, we investigate Bi/BiSb heterojunctions of a finite thickness \cite{Ast2003,Koroteev2004,Hirahara2006,HGuo2011,Nakamura2011,Ohtsubo2013,Benia2015,Ito2016} based on the tight-binding model for bulk Bi and Sb proposed by Liu and Allen \cite{Liu1995}; specifically, we focus on (111) films and heterojunctions because (111) is the most easily grown plane, and thus, the most investigated orientation. The Liu--Allen model provides results that quantitatively agree well with experiments on bulk Bi and Sb \cite{Fu2007,Hasan2010,XLQi2011}. Similar to that of other band calculations \cite{Golin1968,Aguilera2015}, the $Z_2$ topology of pure Bi is trivial in the Liu--Allen model \cite{Fu2007,Teo2008}. Further, the Liu--Allen model can reproduce the band inversion at the $L$-point and the topological transition by extrapolating the parameters for pure Bi and Sb \cite{Teo2008,Fuseya2015b,Fuseya2018}. We adopt a simple virtual crystal approximation with a linear extrapolation of tight-binding parameters of pure Bi and Sb \cite{Fuseya2015b,Fuseya2018}. Within this approximation, the topological transition occurs at $x_c=0.02$ \cite{SeeSM}, which is less than the experimental value $x_c^{\rm exp}\simeq 0.04$ \cite{Lerner1968,Tichovolsky1969,Brandt1972,Oelgart1976,Lenoir1996}. The quantitative mismatch can be attributed to the overly simple linear extrapolation. However, this mismatch does not affect the overall quantum size effect findings. 
The extension of the Liu--Allen model for a finite-thickness film is straightforward \cite{Saito2016,Fuseya2018}. There have previously been concerns that the Liu--Allen model cannot provide results that agree with experiments on the (111) surface \cite{Teo2008}; however, this problem has already been overcome by Saito {\it et al.}, who considered the effects of the surface potential gradient \cite{Saito2016}. In the following calculation, we adopt their surface hopping terms, which can provide results that agree with the experiments on the SS [cf. Fig. \ref{Fig2} (a) and (b)] \cite{Ast2003,Koroteev2004,Hirahara2006,Ohtsubo2013,Benia2015,Ito2016}.
To examine the heterojunction between $n$-bilayer Bi and $n$-bilayer BiSb ($n+n$ BL), we used parameters $z = 1 \to n$ BL for BiSb and $z=n \to 2n$ BL for Bi [Fig. \ref{Fig2} (d)]. 
In the following section, we predominantly provide the results for the $50+50$ BL (20+20 nm) heterojunction; however, we verified that the results were essentially unchanged up to the $100+100$ BL (40+40 nm) system as shown in Sec. \ref{Thickness}.
Additionally, we determined that there was a minimal effect of the interface potential on the following results \cite{SeeSM}.

%%%%%%%%%%%%%%%%%%%%%%%%%%%%%%%%%%%%%%%%%%%%%%%%%%%%%%%%%%%%%%%%%%%%%%%%
\section{Results}
%\subsection{Free standing Bi and BiSb}
%========================================================================
\begin{figure}
 \includegraphics[width=8cm]{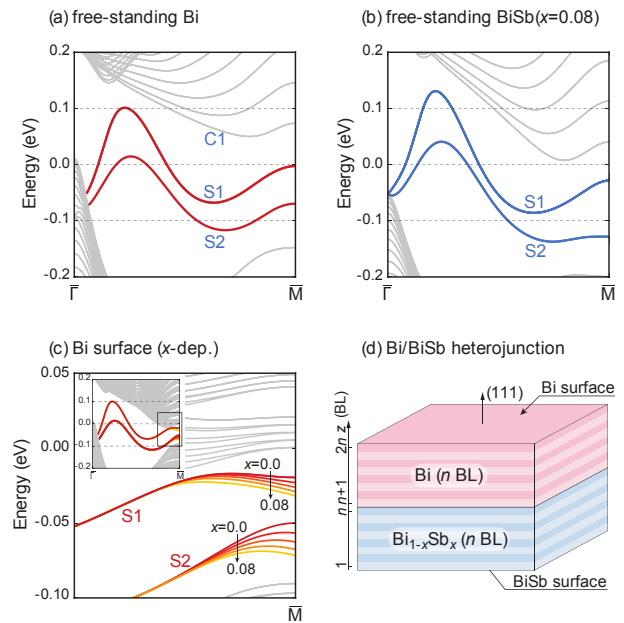}
 \caption{\label{Fig2} Energy dispersion of (a) free-standing Bi (50 BL) and (b) free-standing BiSb($x=0.08$) (50 BL). (c) $x$-dependence of the SS on the Bi surface of Bi/BiSb($x=0.08$) ($50+50$ BL) around the $\bar{M}$-point. The inset shows the $x$-dependence of the entire $\bar{\Gamma}$-$\bar{M}$ region. (d) Schematic of the Bi/BiSb heterojunction ($n+n$ BL). }
 \end{figure}
%========================================================================

Before considering the Bi/BiSb heterojunction, we briefly examined the properties of the (111) surface of free-standing Bi and BiSb. 
The energy dispersions for free-standing trivial Bi (50 BL) and free-standing nontrivial BiSb (50 BL, $x=0.08$: deep inside the nontrivial region \cite{SeeSM}) are plotted in Fig. \ref{Fig2} (a) and (b), respectively. The eigen energies corresponding to the SS are colored in red for Bi and in blue for BiSb. The gray lines correspond to the eigenenergy of the bulk state. 
There are two SSs between the $\bar{\Gamma}$-$\bar{M}$ points. (The $\bar{\Gamma}$ ($\bar{M}$) point in the surface Brillouin zone corresponds to the $T$ ($L$) point in the bulk Brillouin zone, where holes (electrons) are located \cite{SeeSM}.) 
We label the upper energy SS as ``S1" and the lower one as ``S2" \cite{Benia2015}. The lowest CB is labeled ``C1." If the film is infinitely thick, there should be no gap between S1 and S2 at the $\bar{M}$-point for $x<x_c$, whereas the gap opens for $x>x_c$ \cite{Fuseya2018}. Conversely, in the finite-thickness film, the gap ($\sim 72$ meV for free-standing 50 BL Bi) opens because of the interference between the opposite-side surfaces. Consequently, there is no observable difference in the energy dispersion of the SSs between $x<x_c$ and $x>x_c$, even though the bulk CB and VB at the $L$-point are inverted, which is consistent with the analytical solutions for BiSb(111) \cite{Fuseya2018}.

\subsection{Sb-content dependence}
%========================================================================
\begin{figure}
 \includegraphics[width=8cm]{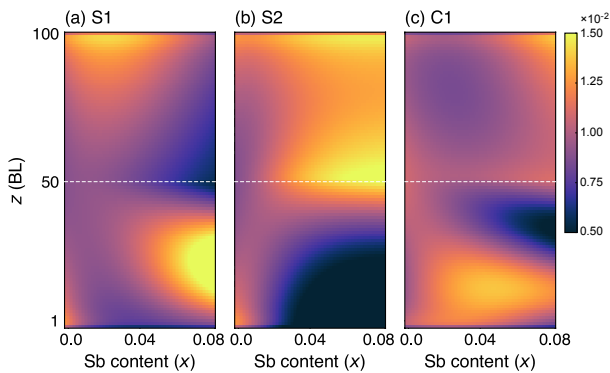}
 \caption{\label{Fig3} Probability distribution $|\psi(z)|^2$ for (a) S1, (b) S2, and (c) C1 in Bi/BiSb ($50+50$ BL). At the BiSb surface ($z=1$ BL) of S1 and S2, the surface state appears for $x\lesssim x_c (\simeq 0.02)$ but does not for $x\gtrsim x_c$, indicating a topological transition.}
 \end{figure}
%========================================================================
Here, we investigate the Sb-content ($x$) dependence of the energy dispersion in the Bi/BiSb heterojunction. Fig. \ref{Fig2} (c) shows the energy dispersion of the SS at the \emph{Bi surface} [Fig. \ref{Fig2} (d)] of the 50+50 BL heterojunction for $x=0$-0.08. Only the parameters in the \emph{BiSb region}, i.e., the blue region in Fig. \ref{Fig2} (d), were varied. Nevertheless, the SS at the Bi surface exhibits an $x$-dependence despite the fact that none of the parameters in the Bi region were changed (the red region in Fig. \ref{Fig2} (d)). This $x$-dependence clearly indicates that the SS at the Bi surface is affected by the BiSb region, though they are separated by more than 50 BL ($\simeq 20$ nm).
It is worth noting that the SS away from the $\bar{M}$-point does not exhibit any $x$-dependence, as shown in the inset of Fig. \ref{Fig2} (c).

Figure \ref{Fig3} shows the $x$-dependence of the probability distribution $|\psi(z)|^2$ for (a) S1, (b) S2, and (c) C1 in Bi/BiSb (50+50 BL). 
The properties of $|\psi (z)|^2$ change from trivial ($x\lesssim x_c$) to nontrivial ($x\gtrsim x_c)$, though the topological transition is smeared due to the finite thickness. 
For $x \lesssim x_c$ (trivial/trivial), 
$|\psi(z)|^2$ of S1 and S2 exhibit peaks (the bright region) at the Bi surface ($z=100$ BL) and the BiSb surface ($z=1$ BL). These peaks are regarded as the SSs of the heterojunction. There is no peak near the interface ($z=50$ BL).
%For $x \lesssim x_c$ (trivial/trivial), the SSs (S1 and S2) both appear at the Bi surface ($z=100$ BL) and the BiSb surface ($z=1$ BL). There is no IS.
Therefore, the system behaves as if it is a homogeneous 100 BL system.
Meanwhile, for $x\gtrsim x_c$ (trivial/nontrivial), 
the peak of $|\psi(z)|^2$ at the BiSb surface disappears, and the peak appears at the interface in S2, i.e., IS appears.
%the SS at the BiSb surface disappears and the IS appears for S2 at approximately $z=50$ BL.
The disappearance of the peak at the BiSb surface for $x>x_c$ is fully consistent with the analytical solution \cite{Fuseya2018}.
The appearance of the IS for $x\gtrsim x_c$ is also consistent with the intuitive picture elucidated in the introduction, and therefore, these results seem reasonable.
However, there is one unexpected property: the IS appears in S2 but not in S1. We will reveal the origin of this unexpected property later.

\subsection{Spatial dependence}

The nature of the SS and IS can be understood more intuitively if we consider the single-particle spectrum $A(\bm{k}_\parallel, z, \varepsilon)$. Contrary to the eigenvalues (Fig. \ref{Fig2}), the spatial ($z$) dependence can be resolved from $A(\bm{k}_\parallel, z, \varepsilon)$. 
Furthermore, $A(\bm{k}_\parallel, z, \varepsilon)$ directly corresponds to angle-resolved photoemission spectroscopy (ARPES) measurements, making the comparison between theory and experiment more straightforward.
The single-particle spectrum is calculated based on the standard definition of the form:
\begin{align}
    A(\bm{k}_\parallel, z, \varepsilon) &= -\frac{1}{\pi}{\rm Tr}\, {\rm Im}\, G(\bm{k}_\parallel, z,  \varepsilon),
    \\
    G(\bm{k}_\parallel, z, \varepsilon) &= \left[ \varepsilon -\mathscr{H}(\bm{k}_\parallel, z) + i\varSigma'' \right]^{-1},
\end{align}
where $G$ is Green's function and $\varSigma''$ is the imaginary part of the self-energy, which is set to $\varSigma''= 0.01$ eV in this paper. (The real part of the self-energy is already included in the Hamiltonian $\mathscr{H}$.) 
The trace is taken over the Hamiltonian of one BL of interest, which consists of 16 bases ($s_{\uparrow \downarrow}$, $p_{x \uparrow \downarrow}$, $p_{y \uparrow \downarrow}$, $p_{z \uparrow \downarrow} \times 2$).
The local spectra of Bi/BiSb($x=0.08$) are plotted in \cite{SeeSM} for both the Bi surface ($z=100$ BL) and BiSb surface ($z=1$ BL), which perfectly match with those obtained from the eigenvalues of $\mathscr{H}$.

%======================================================================== 
 \begin{figure}
 \includegraphics[width=8cm]{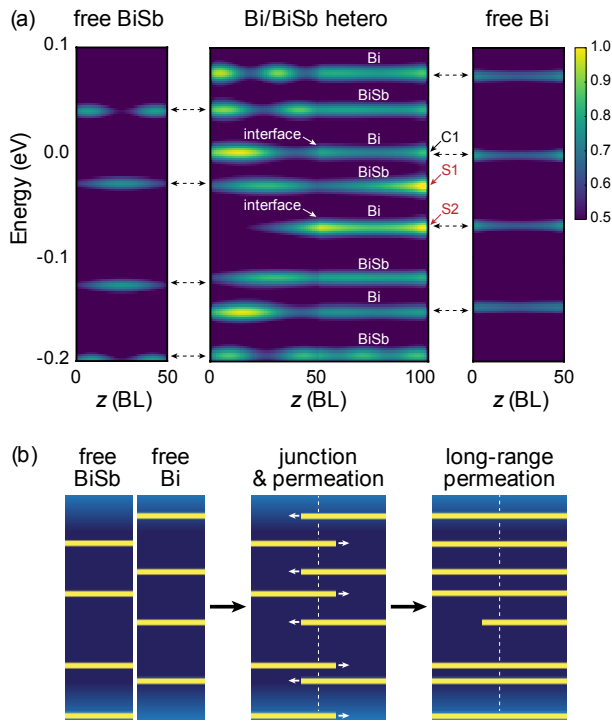}
 \caption{\label{Fig4} (a) Comparison of $A(\bar{M},  z, \varepsilon)$ of the Bi/BiSb heterojunction and free-standing Bi and BiSb. (For the free-standing Bi and BiSb, the magnitude of $A$ is reduced by 1/2 because their thickness is half that of the Bi/BiSb heterojunction.) (b) Schematic of the permeation in the heterojunction.}
 \end{figure}
%========================================================================

The band alignment can be microscopically calculated from the $z$-dependence of the single-particle spectrum, which is plotted in Fig. \ref{Fig02} for (a) $\bm{k}_\parallel=\bar{\Gamma}$ and (d) $\bm{k}_\parallel = \bar{M}$. We observe from Fig. \ref{Fig02} (a) that the band alignment at the $\bar{\Gamma}$-point agrees well with the typical band alignment of the conventional heterojunction, Fig. \ref{Fig0} (a). This is reasonable because there is no band-inversion at the $\bar{\Gamma}$-point (the $T$-point in bulk).
However, the band alignment at the $\bar{M}$-point [Fig. \ref{Fig02} (b)], where the band inversion occurs in bulk, is entirely different from that in Fig. (a) and (b). Each band does not bend around the interface but exhibits flat energy levels without any $z$-dependence.
To clarify the origin of this unexpected property of $A(\bar{M},  z, \varepsilon)$, we compared the $A(\bar{M},  z, \varepsilon)$ of the Bi/BiSb heterojunction with that of free-standing Bi and BiSb, as shown in Fig. \ref{Fig4}, which provided the most significant results of this study.
Notably, we found that each energy level of Bi/BiSb coincides with that of the free-standing Bi and BiSb, while alternating between Bi and BiSb. This correspondence shows that the wave functions of the BiSb region permeate those of the Bi region and vice versa. 
The wave functions of each slab are not connected as is expected from Fig. \ref{Fig0} (b), but they penetrate the opposite side and retain their energy, with no influence from the joined slab. A schematic of the long-range permeation is depicted in Fig. \ref{Fig4} (b).
From this comparison, it is obvious that S1 on the Bi surface originates from the eigenenergy of BiSb and that the wave function of S2 permeates deeply into the BiSb region. Consequently, both S1 and S2 at the Bi surface are affected by increasing $x$, which provides a reasonable interpretation of the $x$-dependence of S1 and S2, as shown in Fig. \ref{Fig2} (c).
Interestingly, the $x$-dependence of $A(\bar{M}, z, \varepsilon)$ on the Bi surface perfectly matches that on the BiSb surface, except for the disappearance of the SS on the BiSb surface, as shown in \cite{SeeSM}. This correspondence indicates the strong correlation between the Bi and BiSb surfaces (their correlation length can be 80 nm \cite{SeeSM}) due to the exceptionally long-range permeation.

The long-range permeation is realized by the long Fermi wavelength ($\sim 100$ nm) of electrons at the $\bar{M}$ point. Meanwhile, the wavelength of holes at the $\bar{\Gamma}$ point is not so long ($\sim 10$ nm). Consequently, no clear permeation is realized at the $\bar{\Gamma}$ point as shown in Fig. \ref{Fig02} (a); this is expected from the ordinary band alignment shown in Fig. \ref{Fig0} (a). The difference in the wavelength originates from the anisotropy of the effective mass: $m^*/m_0=0.00585$ for electrons at the $\bar{M}$ point and $m^*/m_0=0.721$ for holes at the $\bar{\Gamma}$ point \cite{ZZhu2011}.

It is not surprising that the mutual permeation of the wavefunction occurs in the case where the energy levels of one side are almost equal to those of the other, such as $x\lesssim x_c$. In the case where the energy levels are clearly different from each other, such as $x=0.08$, it is naively expected that the energy levels of one side naturally continue to the other by bending the energy level at the interface. However, no such level bending was observed. Instead, the energy levels permeate with each other, maintaining the energy levels constant without bending. This is an unexpected result.

It is worth noting that the previously reported topological ``proximity" effect \cite{GWu2013,Shoman2015,Eremeev2013,Hirahara2017} could be understood as the precursor of this permeating wave function. However, the previously observed permeation is extremely short-range, less than 1.5 nm.
In contrast, our results clearly demonstrate that much longer-range permeation (as long as 40 nm as shown later) of the wave functions occurs in ``thick" topological heterojunctions.

The IS, a distinct intensity at the interface $z=50$ BL, appears not only in S2 but also in C1 (the level just above S1), as shown in Fig. \ref{Fig4} (a), which is consistent with  Fig. \ref{Fig3} (c). Both S2 and C1 originate from the Bi wave function.
As is clear from Fig. \ref{Fig4}, the appearance of the IS cannot be due to what would be expected from Fig. \ref{Fig0} (b). Instead, the IS in S2 and C1 should be interpreted as the SSs of the 50 BL Bi.
Based on this understanding, it is reasonable to infer that the S1 does not have an IS (a distinct intensity at the interface) because it originates from BiSb, which does not have SSs due to the topological restriction for $x>x_c$. This provides the solution to the unpredicted lack of an IS in S1.

From Fig. \ref{Fig4}, we can obtain information regarding another important property of Bi/BiSb. Suppose the scenario of the ARPES of the \emph{Bi surface} with BiSb as a substrate. We will clearly observe two SSs --- S1 and S2 --- in the ARPES spectrum. We may interpret both S1 and S2 as the SSs of Bi. However, S1 does not appear in the free-standing Bi. S1 appears because of the substrate of BiSb. In this sense, S1 is not the true SS of Bi, but is instead the ``superficial" SS originating from the substrate BiSb due to the long-range permeation.

Our scenario of the long-range permeation would be applicable to heterojunctions that have long wavelengths comparable to Bi.
Here, we briefly examine two different types of topological heterojunctions, TlBiTe$_2$/InBiTe$_2$ and Bi$_2$Te$_3$/Bi.
TlBiTe$_2$ has Dirac-like electrons at the $\Gamma$-point \cite{Chen2010,Eremeev2011}. The situation is similar to the $L$-point of Bi. Therefore, the permeation of wavefunction and superficial SS may be observed even in TlBiTe$_2$/InBiTe$_2$. However, the effect is expected to be less significant than Bi/BiSb because the wavelength of TlBiTe$_2$ is much shorter than that of Bi \cite{Paraskevopoulos1985,Chen2010,Eremeev2011}.
The heterojunction of Bi$_2$Te$_3$/Bi indicates another interesting aspect. The large band gap at the $\bar{M}$-point of Bi$_2$Te$_3$ may prevent the Bi electrons from permeating into the Bi$_2$Te$_3$ side \cite{Michiardi2014}. However, according to our observation, the Bi electrons can permeate into the attached substance owing to its long wavelength even if there are no corresponding energy levels. Therefore, a competition between blocking of the Bi$_2$Te$_3$ side and permeation from the Bi side will occur, which would be an interesting future problem.

\subsection{Thickness dependence}\label{Thickness}
%======================================================================== 
 \begin{figure}
 \includegraphics[width=8cm]{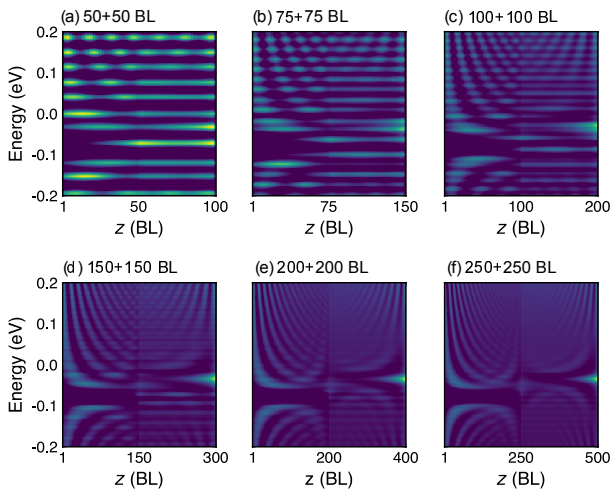}
 \caption{\label{Fig6} Single-particle spectrum $A(\bar{M}, z, \varepsilon)$ of Bi/BiSb($x=0.08$) from (a) $50+50$ BL to (f) $250+250$ BL.}
 \end{figure}
%========================================================================

The single-particle spectra $A(\bm{k}_\parallel, z, \varepsilon)$ at $\bm{k}_\parallel =\bar{M}$ for different thickness from $50+50$ to $250+250$ BL are shown in Fig. \ref{Fig6}.
The long-range permeation, where the eigenenergy is common between Bi ($z> n$) and BiSb region $(z \le n)$ is observed for $50+50$, $75+75$, and $100+100$ BL.
In contrast, the eingenenergy in the Bi region is different from that in the BiSb region for $200+200$ and $250+250$ BL. The long-range permeation is not observed for $n\ge 200$ BL. 
Bi/BiSb of $150+150$ BL is the marginal between the permeated and non-permeated heterojunction.

The spectrum for $250+250$ [Fig. \ref{Fig6} (f)] shows the subtle connection between the bottom CB of BiSb and the top VB of Bi. This connection may be an indication of the band connection as expected in the topological heterojunction, Fig. \ref{Fig0} (b). However, another side connection between the top VB of BiSb and the bottom CB of Bi is not observed. The band cross expected in the topological heterojunction has not been observed even for thick films. 

%======================================================================== 
 \begin{figure}
 \includegraphics[width=8cm]{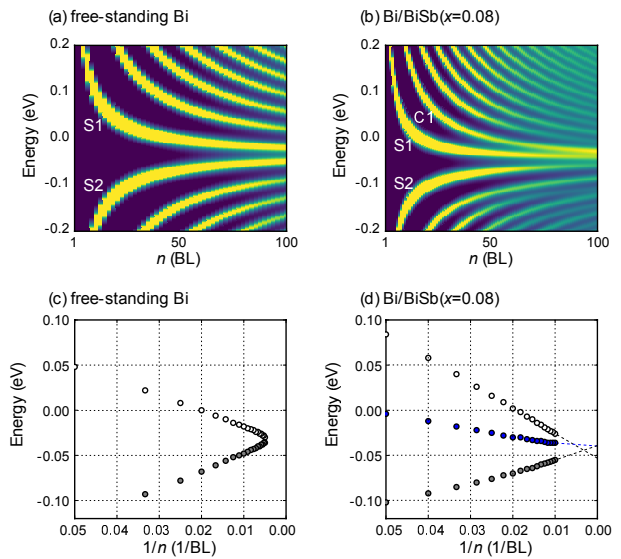}
 \caption{\label{Fig5} Thickness dependence of $A(\bar{M}, \varepsilon)$ at the Bi surface for (a) free-standing Bi ($n$ BL) and (b) Bi/BiSb ($x=0.08$) ($n+n$ BL), where the color bar is similar to that in Fig. \ref{Fig4}. Their peak positions are plotted as a function of $1/n$ for S1, S2, and C1 in (c) free-standing Bi and (d) Bi/BiSb($x=0.08$). The dashed lines in (d) are the extrapolated lines obtained by the least squares method for thick layers ($n\ge 80$ BL). }
 \end{figure}
%========================================================================
%======================================================================== 
 \begin{figure}
 \includegraphics[width=5cm]{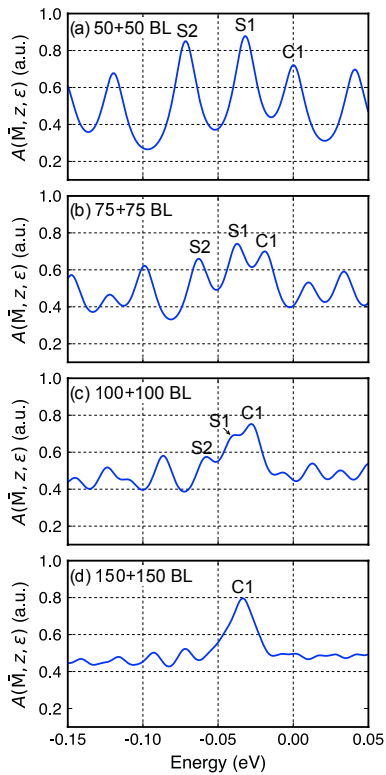}
 \caption{\label{Fig7} Single-particle spectrum at the Bi surface for the $\bar{M}$ point, $A(\bar{M}, 2n, \varepsilon)$, from (a) 50+50 BL to (d) 150+150 BL.
 }
 \end{figure}
%========================================================================

The unusual behavior of S1, the superficial SS, can be more clearly noticed in the thickness dependence. Figure \ref{Fig5} shows the thickness ($n$) dependence of $A(\bar{M}, z, \varepsilon)$ of the Bi side surface at the $\bar{M}$-point for (a) the freestanding Bi and (b) heterojunction Bi/BiSb. The peak positions of S1 and S2 are plotted as a function of $1/n$ in Fig. \ref{Fig5} (c) for the free-standing Bi. The thickness dependences of S1 and S2 are almost symmetric with respect to the center of the band gap for the free-standing Bi. The symmetry between S1 and S2 is a characteristic property of Dirac electrons \cite{Fuseya2018}. Furthermore, it is obvious that the surface gap between S1 and S2 seems to close for $n = 200 \sim 300$ BL reflecting the topologically trivial characteristic.

However, the symmetry between S1 and S2 is broken in the Bi surface of Bi/BiSb. Furthermore, the gap between S1 and S2 is more significant than in the free-standing Bi, and it will never close around $n = 200 \sim 300$ BL. This symmetry breaking and gap opening can be easily understood by considering that S1 originates from the BiSb region. Therefore, we must consider C1 (not S1) to obtain true information regarding the Bi surface. In fact, the thickness dependence of S2 is more symmetric with C1, and the gap between C1 and S2 will close around $n = 200 \sim 300$; both of these findings are consistent with the properties of free-standing Bi.
These results provide an important means for the analysis of the SS of Bi. When one measures the surface states using ARPES, S1 and S2 will exhibit the greatest intensity for thin heterojunction, such as $50+50$ BL. However, evaluation of the gap based on S1 and S2 will lead to a wrong judgement regarding the topology. One should evaluate the gap based on C1 and S2 in order to obtain the true property of the Bi surface, which will provide us with the accurate topological property.
It is worth noting that the asymmetry of the thickness dependence between the two SSs has been observed in pure Bi deposited on a Ge substrate \cite{Ito2016}. Although our results cannot be directly applied to their system because of the different substrate, the asymmetry between the SSs observed in the experiment may be related to the superficial SS affected by the substrate.

Figure \ref{Fig7} shows the detailed spectra of $A(\bar{M}, z, \varepsilon)$ from $50+50$ BL to $150+150$ BL. In thin films of $50+50$ BL, the intensity of S1 is larger than C1. By increasing the thickness, the intensity of S1 reduces, whereas that of C1 enhances. In the $100+100$ BL, the intensity of C1 becomes larger than S1, and the S1 peak is absorbed into the C1 peak. For thick heterojunctions, $n\gtrsim 150$, only the C1 peak survives.
This is another evidence that C1 is the true SS that survives and S1 is the superficial SS that disappears in the bulk limit.

%%%%%%%%%%%%%%%%%%%%%%%%%%%%%%%%%%%%%%%%%%%%%%%%%%%%%%%%%%%%%%%%%%%%%%%%
\section{Summary}
We have investigated the quantum size effects in the heterojunction Bi/BiSb, where the wavelength of the electrons is large. 
We found an unexpected band alignment, whereby the wave functions of the electrons permeate through each other. The wave functions of the two substances coexist throughout the entire heterodevice, even though the substances are spatially separated. 
%Specifically, we can fabricate ``wave-functional alloys" purely by engineering the heterojunction, without the requirement of alloying the substances. This wave-functional alloying technique will open a new avenue for band engineering without introducing a random distribution of atoms.

Additionally, we found that one of the two major spectra on the Bi side surface, S1, was not the true SS of Bi. S1 is the superficial SS originating from the BiSb region, which is far away from the Bi surface. To obtain the true characteristic of the Bi surface, one should consider C1, which is just above S1.
Our results strongly suggest that the SS can be drastically affected by the substrate when the wavelength is large, even when a thick film ($\sim 40$ nm) is used.

% If you have acknowledgments, this puts in the proper section head.
\begin{acknowledgments}
We thank T. Hirahara, L. Perfetti, I. Matsuda, Y. Ohtsubo, A. Takayama, and S. Ito for the fruitful discussions.
This work was supported by the JSPS [Grant No. 19H01850, 22K18318]. 
\end{acknowledgments}

% Create the reference section using BibTeX:
\bibliography{Bismuth.bib,footnote.bib}

%apsrev4-2.bst 2019-01-14 (MD) hand-edited version of apsrev4-1.bst
%Control: key (0)
%Control: author (8) initials jnrlst
%Control: editor formatted (1) identically to author
%Control: production of article title (0) allowed
%Control: page (0) single
%Control: year (1) truncated
%Control: production of eprint (0) enabled
\begin{thebibliography}{52}%
\makeatletter
\providecommand \@ifxundefined [1]{%
 \@ifx{#1\undefined}
}%
\providecommand \@ifnum [1]{%
 \ifnum #1\expandafter \@firstoftwo
 \else \expandafter \@secondoftwo
 \fi
}%
\providecommand \@ifx [1]{%
 \ifx #1\expandafter \@firstoftwo
 \else \expandafter \@secondoftwo
 \fi
}%
\providecommand \natexlab [1]{#1}%
\providecommand \enquote  [1]{``#1''}%
\providecommand \bibnamefont  [1]{#1}%
\providecommand \bibfnamefont [1]{#1}%
\providecommand \citenamefont [1]{#1}%
\providecommand \href@noop [0]{\@secondoftwo}%
\providecommand \href [0]{\begingroup \@sanitize@url \@href}%
\providecommand \@href[1]{\@@startlink{#1}\@@href}%
\providecommand \@@href[1]{\endgroup#1\@@endlink}%
\providecommand \@sanitize@url [0]{\catcode `\\12\catcode `\$12\catcode
  `\&12\catcode `\#12\catcode `\^12\catcode `\_12\catcode `\%12\relax}%
\providecommand \@@startlink[1]{}%
\providecommand \@@endlink[0]{}%
\providecommand \url  [0]{\begingroup\@sanitize@url \@url }%
\providecommand \@url [1]{\endgroup\@href {#1}{\urlprefix }}%
\providecommand \urlprefix  [0]{URL }%
\providecommand \Eprint [0]{\href }%
\providecommand \doibase [0]{https://doi.org/}%
\providecommand \selectlanguage [0]{\@gobble}%
\providecommand \bibinfo  [0]{\@secondoftwo}%
\providecommand \bibfield  [0]{\@secondoftwo}%
\providecommand \translation [1]{[#1]}%
\providecommand \BibitemOpen [0]{}%
\providecommand \bibitemStop [0]{}%
\providecommand \bibitemNoStop [0]{.\EOS\space}%
\providecommand \EOS [0]{\spacefactor3000\relax}%
\providecommand \BibitemShut  [1]{\csname bibitem#1\endcsname}%
\let\auto@bib@innerbib\@empty
%</preamble>
\bibitem [{\citenamefont {Kittel}(2005)}]{Kittel_book_intro}%
  \BibitemOpen
  \bibfield  {author} {\bibinfo {author} {\bibfnamefont {C.}~\bibnamefont
  {Kittel}},\ }\href@noop {} {\emph {\bibinfo {title} {Introduction to Solid
  State Physics}}}\ (\bibinfo  {publisher} {Wiley},\ \bibinfo {year}
  {2005})\BibitemShut {NoStop}%
\bibitem [{\citenamefont {Ihn}(2010)}]{Ihn_book}%
  \BibitemOpen
  \bibfield  {author} {\bibinfo {author} {\bibfnamefont {T.}~\bibnamefont
  {Ihn}},\ }\href {https://books.google.co.jp/books?id=UdgVDAAAQBAJ} {\emph
  {\bibinfo {title} {Semiconductor Nanostructures: Quantum States and
  Electronic Transport}}}\ (\bibinfo  {publisher} {OUP Oxford},\ \bibinfo
  {year} {2010})\BibitemShut {NoStop}%
\bibitem [{\citenamefont {Bastard}\ \emph {et~al.}(2017)\citenamefont
  {Bastard}, \citenamefont {Carosella},\ and\ \citenamefont
  {Ndebeka-bandou}}]{Bastard_book}%
  \BibitemOpen
  \bibfield  {author} {\bibinfo {author} {\bibfnamefont {G.}~\bibnamefont
  {Bastard}}, \bibinfo {author} {\bibfnamefont {F.}~\bibnamefont {Carosella}},\
  and\ \bibinfo {author} {\bibfnamefont {C.}~\bibnamefont {Ndebeka-bandou}},\
  }\href {https://books.google.co.jp/books?id=rM02DwAAQBAJ} {\emph {\bibinfo
  {title} {Quantum States And Scattering In Semiconductor Nanostructures}}},\
  Advanced Textbooks In Physics\ (\bibinfo  {publisher} {World Scientific
  Publishing Company},\ \bibinfo {year} {2017})\BibitemShut {NoStop}%
\bibitem [{\citenamefont {Volkov}\ and\ \citenamefont
  {Pankratov}(1985)}]{Volkov1985}%
  \BibitemOpen
  \bibfield  {author} {\bibinfo {author} {\bibfnamefont {B.~A.}\ \bibnamefont
  {Volkov}}\ and\ \bibinfo {author} {\bibfnamefont {O.~A.}\ \bibnamefont
  {Pankratov}},\ }\bibfield  {title} {\bibinfo {title} {Two-dimensional
  massless electrons in an inverted contact},\ }\href@noop {} {\bibfield
  {journal} {\bibinfo  {journal} {JETP Lett}\ }\textbf {\bibinfo {volume}
  {42}},\ \bibinfo {pages} {178} (\bibinfo {year} {1985})}\BibitemShut
  {NoStop}%
\bibitem [{\citenamefont {Fradkin}\ \emph {et~al.}(1986)\citenamefont
  {Fradkin}, \citenamefont {Dagotto},\ and\ \citenamefont
  {Boyanovsky}}]{Fradkin1986}%
  \BibitemOpen
  \bibfield  {author} {\bibinfo {author} {\bibfnamefont {E.}~\bibnamefont
  {Fradkin}}, \bibinfo {author} {\bibfnamefont {E.}~\bibnamefont {Dagotto}},\
  and\ \bibinfo {author} {\bibfnamefont {D.}~\bibnamefont {Boyanovsky}},\
  }\bibfield  {title} {\bibinfo {title} {Physical realization of the parity
  anomaly in condensed matter physics},\ }\href
  {https://doi.org/10.1103/PhysRevLett.57.2967} {\bibfield  {journal} {\bibinfo
   {journal} {Phys. Rev. Lett.}\ }\textbf {\bibinfo {volume} {57}},\ \bibinfo
  {pages} {2967} (\bibinfo {year} {1986})}\BibitemShut {NoStop}%
\bibitem [{\citenamefont {Cade}(1985)}]{Cade1985}%
  \BibitemOpen
  \bibfield  {author} {\bibinfo {author} {\bibfnamefont {N.~A.}\ \bibnamefont
  {Cade}},\ }\bibfield  {title} {\bibinfo {title} {Quantum well bound states of
  {HgTe} in {CdTe}},\ }\href {https://doi.org/10.1088/0022-3719/18/26/024}
  {\bibfield  {journal} {\bibinfo  {journal} {Journal of Physics C: Solid State
  Physics}\ }\textbf {\bibinfo {volume} {18}},\ \bibinfo {pages} {5135}
  (\bibinfo {year} {1985})}\BibitemShut {NoStop}%
\bibitem [{\citenamefont {Chang}\ \emph {et~al.}(1985)\citenamefont {Chang},
  \citenamefont {Schulman}, \citenamefont {Bastard}, \citenamefont {Guldner},\
  and\ \citenamefont {Voos}}]{Chang1985}%
  \BibitemOpen
  \bibfield  {author} {\bibinfo {author} {\bibfnamefont {Y.-C.}\ \bibnamefont
  {Chang}}, \bibinfo {author} {\bibfnamefont {J.~N.}\ \bibnamefont {Schulman}},
  \bibinfo {author} {\bibfnamefont {G.}~\bibnamefont {Bastard}}, \bibinfo
  {author} {\bibfnamefont {Y.}~\bibnamefont {Guldner}},\ and\ \bibinfo {author}
  {\bibfnamefont {M.}~\bibnamefont {Voos}},\ }\bibfield  {title} {\bibinfo
  {title} {Effects of quasi-interface states in hgte-cdte superlattices},\
  }\href {https://doi.org/10.1103/PhysRevB.31.2557} {\bibfield  {journal}
  {\bibinfo  {journal} {Phys. Rev. B}\ }\textbf {\bibinfo {volume} {31}},\
  \bibinfo {pages} {2557} (\bibinfo {year} {1985})}\BibitemShut {NoStop}%
\bibitem [{\citenamefont {Lin-Liu}\ and\ \citenamefont
  {Sham}(1985)}]{LinLiu1985}%
  \BibitemOpen
  \bibfield  {author} {\bibinfo {author} {\bibfnamefont {Y.~R.}\ \bibnamefont
  {Lin-Liu}}\ and\ \bibinfo {author} {\bibfnamefont {L.~J.}\ \bibnamefont
  {Sham}},\ }\bibfield  {title} {\bibinfo {title} {Interface states and
  subbands in hgte-cdte heterostructures},\ }\href
  {https://doi.org/10.1103/PhysRevB.32.5561} {\bibfield  {journal} {\bibinfo
  {journal} {Phys. Rev. B}\ }\textbf {\bibinfo {volume} {32}},\ \bibinfo
  {pages} {5561} (\bibinfo {year} {1985})}\BibitemShut {NoStop}%
\bibitem [{\citenamefont {Pankratov}\ \emph {et~al.}(1987)\citenamefont
  {Pankratov}, \citenamefont {Pakhomov},\ and\ \citenamefont
  {Volkov}}]{Pankratov1987}%
  \BibitemOpen
  \bibfield  {author} {\bibinfo {author} {\bibfnamefont {O.}~\bibnamefont
  {Pankratov}}, \bibinfo {author} {\bibfnamefont {S.}~\bibnamefont
  {Pakhomov}},\ and\ \bibinfo {author} {\bibfnamefont {B.}~\bibnamefont
  {Volkov}},\ }\bibfield  {title} {\bibinfo {title} {Supersymmetry in
  heterojunctions: Band-inverting contact on the basis of pb1-xsnxte and
  hg1-xcdxte},\ }\href
  {https://doi.org/https://doi.org/10.1016/0038-1098(87)90934-3} {\bibfield
  {journal} {\bibinfo  {journal} {Solid State Communications}\ }\textbf
  {\bibinfo {volume} {61}},\ \bibinfo {pages} {93} (\bibinfo {year}
  {1987})}\BibitemShut {NoStop}%
\bibitem [{\citenamefont {Hasan}\ and\ \citenamefont {Kane}(2010)}]{Hasan2010}%
  \BibitemOpen
  \bibfield  {author} {\bibinfo {author} {\bibfnamefont {M.~Z.}\ \bibnamefont
  {Hasan}}\ and\ \bibinfo {author} {\bibfnamefont {C.~L.}\ \bibnamefont
  {Kane}},\ }\bibfield  {title} {\bibinfo {title} {\textit{Colloquium} :
  Topological insulators},\ }\href {https://doi.org/10.1103/RevModPhys.82.3045}
  {\bibfield  {journal} {\bibinfo  {journal} {Rev. Mod. Phys.}\ }\textbf
  {\bibinfo {volume} {82}},\ \bibinfo {pages} {3045} (\bibinfo {year}
  {2010})}\BibitemShut {NoStop}%
\bibitem [{\citenamefont {Qi}\ and\ \citenamefont {Zhang}(2011)}]{XLQi2011}%
  \BibitemOpen
  \bibfield  {author} {\bibinfo {author} {\bibfnamefont {X.-L.}\ \bibnamefont
  {Qi}}\ and\ \bibinfo {author} {\bibfnamefont {S.-C.}\ \bibnamefont {Zhang}},\
  }\bibfield  {title} {\bibinfo {title} {Topological insulators and
  superconductors},\ }\href {https://doi.org/10.1103/RevModPhys.83.1057}
  {\bibfield  {journal} {\bibinfo  {journal} {Rev. Mod. Phys.}\ }\textbf
  {\bibinfo {volume} {83}},\ \bibinfo {pages} {1057} (\bibinfo {year}
  {2011})}\BibitemShut {NoStop}%
\bibitem [{\citenamefont {Bernevig}\ \emph {et~al.}(2006)\citenamefont
  {Bernevig}, \citenamefont {Hughes},\ and\ \citenamefont
  {Zhang}}]{Bernevig2006}%
  \BibitemOpen
  \bibfield  {author} {\bibinfo {author} {\bibfnamefont {B.~A.}\ \bibnamefont
  {Bernevig}}, \bibinfo {author} {\bibfnamefont {T.~L.}\ \bibnamefont
  {Hughes}},\ and\ \bibinfo {author} {\bibfnamefont {S.-C.}\ \bibnamefont
  {Zhang}},\ }\bibfield  {title} {\bibinfo {title} {Quantum spin hall effect
  and topological phase transition in hgte quantum wells},\ }\href
  {https://doi.org/10.1126/science.1133734} {\bibfield  {journal} {\bibinfo
  {journal} {Science}\ }\textbf {\bibinfo {volume} {314}},\ \bibinfo {pages}
  {1757} (\bibinfo {year} {2006})}\BibitemShut {NoStop}%
\bibitem [{\citenamefont {Hsieh}\ \emph {et~al.}(2012)\citenamefont {Hsieh},
  \citenamefont {Lin}, \citenamefont {Liu}, \citenamefont {Duan}, \citenamefont
  {Bansil},\ and\ \citenamefont {Fu}}]{Hsieh2012}%
  \BibitemOpen
  \bibfield  {author} {\bibinfo {author} {\bibfnamefont {T.~H.}\ \bibnamefont
  {Hsieh}}, \bibinfo {author} {\bibfnamefont {H.}~\bibnamefont {Lin}}, \bibinfo
  {author} {\bibfnamefont {J.}~\bibnamefont {Liu}}, \bibinfo {author}
  {\bibfnamefont {W.}~\bibnamefont {Duan}}, \bibinfo {author} {\bibfnamefont
  {A.}~\bibnamefont {Bansil}},\ and\ \bibinfo {author} {\bibfnamefont
  {L.}~\bibnamefont {Fu}},\ }\bibfield  {title} {\bibinfo {title} {Topological
  crystalline insulators in the snte material class},\ }\href
  {https://doi.org/10.1038/ncomms1969} {\bibfield  {journal} {\bibinfo
  {journal} {Nature Communications}\ }\textbf {\bibinfo {volume} {3}},\
  \bibinfo {pages} {982} (\bibinfo {year} {2012})}\BibitemShut {NoStop}%
\bibitem [{\citenamefont {Wu}\ \emph {et~al.}(2013)\citenamefont {Wu},
  \citenamefont {Chen}, \citenamefont {Sun}, \citenamefont {Li}, \citenamefont
  {Cui}, \citenamefont {Franchini}, \citenamefont {Wang}, \citenamefont
  {Chen},\ and\ \citenamefont {Zhang}}]{GWu2013}%
  \BibitemOpen
  \bibfield  {author} {\bibinfo {author} {\bibfnamefont {G.}~\bibnamefont
  {Wu}}, \bibinfo {author} {\bibfnamefont {H.}~\bibnamefont {Chen}}, \bibinfo
  {author} {\bibfnamefont {Y.}~\bibnamefont {Sun}}, \bibinfo {author}
  {\bibfnamefont {X.}~\bibnamefont {Li}}, \bibinfo {author} {\bibfnamefont
  {P.}~\bibnamefont {Cui}}, \bibinfo {author} {\bibfnamefont {C.}~\bibnamefont
  {Franchini}}, \bibinfo {author} {\bibfnamefont {J.}~\bibnamefont {Wang}},
  \bibinfo {author} {\bibfnamefont {X.-Q.}\ \bibnamefont {Chen}},\ and\
  \bibinfo {author} {\bibfnamefont {Z.}~\bibnamefont {Zhang}},\ }\bibfield
  {title} {\bibinfo {title} {Tuning the vertical location of helical surface
  states in topological insulator heterostructures via dual-proximity
  effects},\ }\href {https://doi.org/10.1038/srep01233} {\bibfield  {journal}
  {\bibinfo  {journal} {Scientific Reports}\ }\textbf {\bibinfo {volume} {3}},\
  \bibinfo {pages} {1233} (\bibinfo {year} {2013})}\BibitemShut {NoStop}%
\bibitem [{\citenamefont {Shoman}\ \emph {et~al.}(2015)\citenamefont {Shoman},
  \citenamefont {Takayama}, \citenamefont {Sato}, \citenamefont {Souma},
  \citenamefont {Takahashi}, \citenamefont {Oguchi}, \citenamefont {Segawa},\
  and\ \citenamefont {Ando}}]{Shoman2015}%
  \BibitemOpen
  \bibfield  {author} {\bibinfo {author} {\bibfnamefont {T.}~\bibnamefont
  {Shoman}}, \bibinfo {author} {\bibfnamefont {A.}~\bibnamefont {Takayama}},
  \bibinfo {author} {\bibfnamefont {T.}~\bibnamefont {Sato}}, \bibinfo {author}
  {\bibfnamefont {S.}~\bibnamefont {Souma}}, \bibinfo {author} {\bibfnamefont
  {T.}~\bibnamefont {Takahashi}}, \bibinfo {author} {\bibfnamefont
  {T.}~\bibnamefont {Oguchi}}, \bibinfo {author} {\bibfnamefont
  {K.}~\bibnamefont {Segawa}},\ and\ \bibinfo {author} {\bibfnamefont
  {Y.}~\bibnamefont {Ando}},\ }\bibfield  {title} {\bibinfo {title}
  {Topological proximity effect in a topological insulator hybrid},\ }\href
  {https://doi.org/10.1038/ncomms7547} {\bibfield  {journal} {\bibinfo
  {journal} {Nature Communications}\ }\textbf {\bibinfo {volume} {6}},\
  \bibinfo {pages} {6547} (\bibinfo {year} {2015})}\BibitemShut {NoStop}%
\bibitem [{\citenamefont {Krizman}\ \emph {et~al.}(2022)\citenamefont
  {Krizman}, \citenamefont {Assaf}, \citenamefont {Orlita}, \citenamefont
  {Bauer}, \citenamefont {Springholz}, \citenamefont {Ferreira}, \citenamefont
  {de~Vaulchier},\ and\ \citenamefont {Guldner}}]{Krizman2022}%
  \BibitemOpen
  \bibfield  {author} {\bibinfo {author} {\bibfnamefont {G.}~\bibnamefont
  {Krizman}}, \bibinfo {author} {\bibfnamefont {B.~A.}\ \bibnamefont {Assaf}},
  \bibinfo {author} {\bibfnamefont {M.}~\bibnamefont {Orlita}}, \bibinfo
  {author} {\bibfnamefont {G.}~\bibnamefont {Bauer}}, \bibinfo {author}
  {\bibfnamefont {G.}~\bibnamefont {Springholz}}, \bibinfo {author}
  {\bibfnamefont {R.}~\bibnamefont {Ferreira}}, \bibinfo {author}
  {\bibfnamefont {L.~A.}\ \bibnamefont {de~Vaulchier}},\ and\ \bibinfo {author}
  {\bibfnamefont {Y.}~\bibnamefont {Guldner}},\ }\bibfield  {title} {\bibinfo
  {title} {Interaction between interface and massive states in multivalley
  topological heterostructures},\ }\href
  {https://doi.org/10.1103/PhysRevResearch.4.013179} {\bibfield  {journal}
  {\bibinfo  {journal} {Phys. Rev. Research}\ }\textbf {\bibinfo {volume}
  {4}},\ \bibinfo {pages} {013179} (\bibinfo {year} {2022})}\BibitemShut
  {NoStop}%
\bibitem [{\citenamefont {Zhou}\ \emph {et~al.}(2008)\citenamefont {Zhou},
  \citenamefont {Lu}, \citenamefont {Chu}, \citenamefont {Shen},\ and\
  \citenamefont {Niu}}]{BZhou2008}%
  \BibitemOpen
  \bibfield  {author} {\bibinfo {author} {\bibfnamefont {B.}~\bibnamefont
  {Zhou}}, \bibinfo {author} {\bibfnamefont {H.-Z.}\ \bibnamefont {Lu}},
  \bibinfo {author} {\bibfnamefont {R.-L.}\ \bibnamefont {Chu}}, \bibinfo
  {author} {\bibfnamefont {S.-Q.}\ \bibnamefont {Shen}},\ and\ \bibinfo
  {author} {\bibfnamefont {Q.}~\bibnamefont {Niu}},\ }\bibfield  {title}
  {\bibinfo {title} {Finite size effects on helical edge states in a quantum
  spin-hall system},\ }\href {https://doi.org/10.1103/PhysRevLett.101.246807}
  {\bibfield  {journal} {\bibinfo  {journal} {Phys. Rev. Lett.}\ }\textbf
  {\bibinfo {volume} {101}},\ \bibinfo {pages} {246807} (\bibinfo {year}
  {2008})}\BibitemShut {NoStop}%
\bibitem [{\citenamefont {Linder}\ \emph {et~al.}(2009)\citenamefont {Linder},
  \citenamefont {Yokoyama},\ and\ \citenamefont {Sudb\o{}}}]{Linder2009}%
  \BibitemOpen
  \bibfield  {author} {\bibinfo {author} {\bibfnamefont {J.}~\bibnamefont
  {Linder}}, \bibinfo {author} {\bibfnamefont {T.}~\bibnamefont {Yokoyama}},\
  and\ \bibinfo {author} {\bibfnamefont {A.}~\bibnamefont {Sudb\o{}}},\
  }\bibfield  {title} {\bibinfo {title} {Anomalous finite size effects on
  surface states in the topological insulator
  ${\text{bi}}_{2}{\text{se}}_{3}$},\ }\href
  {https://doi.org/10.1103/PhysRevB.80.205401} {\bibfield  {journal} {\bibinfo
  {journal} {Phys. Rev. B}\ }\textbf {\bibinfo {volume} {80}},\ \bibinfo
  {pages} {205401} (\bibinfo {year} {2009})}\BibitemShut {NoStop}%
\bibitem [{\citenamefont {Lu}\ \emph {et~al.}(2010)\citenamefont {Lu},
  \citenamefont {Shan}, \citenamefont {Yao}, \citenamefont {Niu},\ and\
  \citenamefont {Shen}}]{HZLu2010}%
  \BibitemOpen
  \bibfield  {author} {\bibinfo {author} {\bibfnamefont {H.-Z.}\ \bibnamefont
  {Lu}}, \bibinfo {author} {\bibfnamefont {W.-Y.}\ \bibnamefont {Shan}},
  \bibinfo {author} {\bibfnamefont {W.}~\bibnamefont {Yao}}, \bibinfo {author}
  {\bibfnamefont {Q.}~\bibnamefont {Niu}},\ and\ \bibinfo {author}
  {\bibfnamefont {S.-Q.}\ \bibnamefont {Shen}},\ }\bibfield  {title} {\bibinfo
  {title} {Massive dirac fermions and spin physics in an ultrathin film of
  topological insulator},\ }\href {https://doi.org/10.1103/PhysRevB.81.115407}
  {\bibfield  {journal} {\bibinfo  {journal} {Phys. Rev. B}\ }\textbf {\bibinfo
  {volume} {81}},\ \bibinfo {pages} {115407} (\bibinfo {year}
  {2010})}\BibitemShut {NoStop}%
\bibitem [{\citenamefont {Shen}(2012)}]{SQShen_book}%
  \BibitemOpen
  \bibfield  {author} {\bibinfo {author} {\bibfnamefont {S.-Q.}\ \bibnamefont
  {Shen}},\ }\href@noop {} {\emph {\bibinfo {title} {Topological Insulators}}}\
  (\bibinfo  {publisher} {Springer-Verlag Berlin Heidelberg},\ \bibinfo {year}
  {2012})\BibitemShut {NoStop}%
\bibitem [{\citenamefont {Fuseya}\ and\ \citenamefont
  {Fukuyama}(2018)}]{Fuseya2018}%
  \BibitemOpen
  \bibfield  {author} {\bibinfo {author} {\bibfnamefont {Y.}~\bibnamefont
  {Fuseya}}\ and\ \bibinfo {author} {\bibfnamefont {H.}~\bibnamefont
  {Fukuyama}},\ }\bibfield  {title} {\bibinfo {title} {Analytical solutions for
  the surface states of bi$_{1-x}$sb$_x$},\ }\href
  {https://doi.org/10.7566/JPSJ.87.044710} {\bibfield  {journal} {\bibinfo
  {journal} {J. Phys. Soc. Jpn.}\ }\textbf {\bibinfo {volume} {87}},\ \bibinfo
  {pages} {044710} (\bibinfo {year} {2018})}\BibitemShut {NoStop}%
\bibitem [{\citenamefont {Aguilera}\ \emph {et~al.}(2021)\citenamefont
  {Aguilera}, \citenamefont {Kim}, \citenamefont {Friedrich}, \citenamefont
  {Bihlmayer},\ and\ \citenamefont {Bl\"ugel}}]{Aguilera2021}%
  \BibitemOpen
  \bibfield  {author} {\bibinfo {author} {\bibfnamefont {I.}~\bibnamefont
  {Aguilera}}, \bibinfo {author} {\bibfnamefont {H.-J.}\ \bibnamefont {Kim}},
  \bibinfo {author} {\bibfnamefont {C.}~\bibnamefont {Friedrich}}, \bibinfo
  {author} {\bibfnamefont {G.}~\bibnamefont {Bihlmayer}},\ and\ \bibinfo
  {author} {\bibfnamefont {S.}~\bibnamefont {Bl\"ugel}},\ }\bibfield  {title}
  {\bibinfo {title} {$z_2$ topology of bismuth},\ }\href
  {https://doi.org/10.1103/PhysRevMaterials.5.L091201} {\bibfield  {journal}
  {\bibinfo  {journal} {Phys. Rev. Materials}\ }\textbf {\bibinfo {volume}
  {5}},\ \bibinfo {pages} {L091201} (\bibinfo {year} {2021})}\BibitemShut
  {NoStop}%
\bibitem [{\citenamefont {Eremeev}\ \emph {et~al.}(2013)\citenamefont
  {Eremeev}, \citenamefont {Men'shov}, \citenamefont {Tugushev}, \citenamefont
  {Echenique},\ and\ \citenamefont {Chulkov}}]{Eremeev2013}%
  \BibitemOpen
  \bibfield  {author} {\bibinfo {author} {\bibfnamefont {S.~V.}\ \bibnamefont
  {Eremeev}}, \bibinfo {author} {\bibfnamefont {V.~N.}\ \bibnamefont
  {Men'shov}}, \bibinfo {author} {\bibfnamefont {V.~V.}\ \bibnamefont
  {Tugushev}}, \bibinfo {author} {\bibfnamefont {P.~M.}\ \bibnamefont
  {Echenique}},\ and\ \bibinfo {author} {\bibfnamefont {E.~V.}\ \bibnamefont
  {Chulkov}},\ }\bibfield  {title} {\bibinfo {title} {Magnetic proximity effect
  at the three-dimensional topological insulator/magnetic insulator
  interface},\ }\href {https://doi.org/10.1103/PhysRevB.88.144430} {\bibfield
  {journal} {\bibinfo  {journal} {Phys. Rev. B}\ }\textbf {\bibinfo {volume}
  {88}},\ \bibinfo {pages} {144430} (\bibinfo {year} {2013})}\BibitemShut
  {NoStop}%
\bibitem [{\citenamefont {Hirahara}\ \emph {et~al.}(2017)\citenamefont
  {Hirahara}, \citenamefont {Eremeev}, \citenamefont {Shirasawa}, \citenamefont
  {Okuyama}, \citenamefont {Kubo}, \citenamefont {Nakanishi}, \citenamefont
  {Akiyama}, \citenamefont {Takayama}, \citenamefont {Hajiri}, \citenamefont
  {Ideta}, \citenamefont {Matsunami}, \citenamefont {Sumida}, \citenamefont
  {Miyamoto}, \citenamefont {Takagi}, \citenamefont {Tanaka}, \citenamefont
  {Okuda}, \citenamefont {Yokoyama}, \citenamefont {Kimura}, \citenamefont
  {Hasegawa},\ and\ \citenamefont {Chulkov}}]{Hirahara2017}%
  \BibitemOpen
  \bibfield  {author} {\bibinfo {author} {\bibfnamefont {T.}~\bibnamefont
  {Hirahara}}, \bibinfo {author} {\bibfnamefont {S.~V.}\ \bibnamefont
  {Eremeev}}, \bibinfo {author} {\bibfnamefont {T.}~\bibnamefont {Shirasawa}},
  \bibinfo {author} {\bibfnamefont {Y.}~\bibnamefont {Okuyama}}, \bibinfo
  {author} {\bibfnamefont {T.}~\bibnamefont {Kubo}}, \bibinfo {author}
  {\bibfnamefont {R.}~\bibnamefont {Nakanishi}}, \bibinfo {author}
  {\bibfnamefont {R.}~\bibnamefont {Akiyama}}, \bibinfo {author} {\bibfnamefont
  {A.}~\bibnamefont {Takayama}}, \bibinfo {author} {\bibfnamefont
  {T.}~\bibnamefont {Hajiri}}, \bibinfo {author} {\bibfnamefont {S.-i.}\
  \bibnamefont {Ideta}}, \bibinfo {author} {\bibfnamefont {M.}~\bibnamefont
  {Matsunami}}, \bibinfo {author} {\bibfnamefont {K.}~\bibnamefont {Sumida}},
  \bibinfo {author} {\bibfnamefont {K.}~\bibnamefont {Miyamoto}}, \bibinfo
  {author} {\bibfnamefont {Y.}~\bibnamefont {Takagi}}, \bibinfo {author}
  {\bibfnamefont {K.}~\bibnamefont {Tanaka}}, \bibinfo {author} {\bibfnamefont
  {T.}~\bibnamefont {Okuda}}, \bibinfo {author} {\bibfnamefont
  {T.}~\bibnamefont {Yokoyama}}, \bibinfo {author} {\bibfnamefont {S.-i.}\
  \bibnamefont {Kimura}}, \bibinfo {author} {\bibfnamefont {S.}~\bibnamefont
  {Hasegawa}},\ and\ \bibinfo {author} {\bibfnamefont {E.~V.}\ \bibnamefont
  {Chulkov}},\ }\bibfield  {title} {\bibinfo {title} {Large-gap magnetic
  topological heterostructure formed by subsurface incorporation of a
  ferromagnetic layer},\ }\href {https://doi.org/10.1021/acs.nanolett.7b00560}
  {\bibfield  {journal} {\bibinfo  {journal} {Nano Letters}\ }\textbf {\bibinfo
  {volume} {17}},\ \bibinfo {pages} {3493} (\bibinfo {year}
  {2017})}\BibitemShut {NoStop}%
\bibitem [{\citenamefont {Burkov}\ and\ \citenamefont
  {Balents}(2011)}]{Burkov2011}%
  \BibitemOpen
  \bibfield  {author} {\bibinfo {author} {\bibfnamefont {A.~A.}\ \bibnamefont
  {Burkov}}\ and\ \bibinfo {author} {\bibfnamefont {L.}~\bibnamefont
  {Balents}},\ }\bibfield  {title} {\bibinfo {title} {Weyl semimetal in a
  topological insulator multilayer},\ }\href
  {https://doi.org/10.1103/PhysRevLett.107.127205} {\bibfield  {journal}
  {\bibinfo  {journal} {Phys. Rev. Lett.}\ }\textbf {\bibinfo {volume} {107}},\
  \bibinfo {pages} {127205} (\bibinfo {year} {2011})}\BibitemShut {NoStop}%
\bibitem [{\citenamefont {Liu}\ and\ \citenamefont {Allen}(1995)}]{Liu1995}%
  \BibitemOpen
  \bibfield  {author} {\bibinfo {author} {\bibfnamefont {Y.}~\bibnamefont
  {Liu}}\ and\ \bibinfo {author} {\bibfnamefont {R.~E.}\ \bibnamefont
  {Allen}},\ }\bibfield  {title} {\bibinfo {title} {Electronic structure of the
  semimetals bi and sb},\ }\href@noop {} {\bibfield  {journal} {\bibinfo
  {journal} {Phys. Rev. B}\ }\textbf {\bibinfo {volume} {52}},\ \bibinfo
  {pages} {1566} (\bibinfo {year} {1995})}\BibitemShut {NoStop}%
\bibitem [{\citenamefont {Zhu}\ \emph {et~al.}(2011)\citenamefont {Zhu},
  \citenamefont {Fauqu\'e}, \citenamefont {Fuseya},\ and\ \citenamefont
  {Behnia}}]{ZZhu2011}%
  \BibitemOpen
  \bibfield  {author} {\bibinfo {author} {\bibfnamefont {Z.}~\bibnamefont
  {Zhu}}, \bibinfo {author} {\bibfnamefont {B.}~\bibnamefont {Fauqu\'e}},
  \bibinfo {author} {\bibfnamefont {Y.}~\bibnamefont {Fuseya}},\ and\ \bibinfo
  {author} {\bibfnamefont {K.}~\bibnamefont {Behnia}},\ }\bibfield  {title}
  {\bibinfo {title} {Angle resolved landau spectrum of electrons and holes in
  bismuth},\ }\href@noop {} {\bibfield  {journal} {\bibinfo  {journal} {Phys.
  Rev. B}\ }\textbf {\bibinfo {volume} {84}},\ \bibinfo {pages} {115137}
  (\bibinfo {year} {2011})}\BibitemShut {NoStop}%
\bibitem [{\citenamefont {Fuseya}\ \emph
  {et~al.}(2015{\natexlab{a}})\citenamefont {Fuseya}, \citenamefont {Ogata},\
  and\ \citenamefont {Fukuyama}}]{Fuseya2015}%
  \BibitemOpen
  \bibfield  {author} {\bibinfo {author} {\bibfnamefont {Y.}~\bibnamefont
  {Fuseya}}, \bibinfo {author} {\bibfnamefont {M.}~\bibnamefont {Ogata}},\ and\
  \bibinfo {author} {\bibfnamefont {H.}~\bibnamefont {Fukuyama}},\ }\bibfield
  {title} {\bibinfo {title} {Transport properties and diamagnetism of dirac
  electrons in bismuth},\ }\href@noop {} {\bibfield  {journal} {\bibinfo
  {journal} {J. Phys. Soc. Jpn.}\ }\textbf {\bibinfo {volume} {84}},\ \bibinfo
  {pages} {012001} (\bibinfo {year} {2015}{\natexlab{a}})}\BibitemShut
  {NoStop}%
\bibitem [{\citenamefont {Lerner}\ \emph {et~al.}(1968)\citenamefont {Lerner},
  \citenamefont {Cuff},\ and\ \citenamefont {Williams}}]{Lerner1968}%
  \BibitemOpen
  \bibfield  {author} {\bibinfo {author} {\bibfnamefont {L.~S.}\ \bibnamefont
  {Lerner}}, \bibinfo {author} {\bibfnamefont {K.~F.}\ \bibnamefont {Cuff}},\
  and\ \bibinfo {author} {\bibfnamefont {L.~R.}\ \bibnamefont {Williams}},\
  }\bibfield  {title} {\bibinfo {title} {Energy-band parameters and relative
  band-edge motions in the bi-sb alloy system near the
  semimetal\char22{}semiconductor transition},\ }\href
  {https://doi.org/10.1103/RevModPhys.40.770} {\bibfield  {journal} {\bibinfo
  {journal} {Rev. Mod. Phys.}\ }\textbf {\bibinfo {volume} {40}},\ \bibinfo
  {pages} {770} (\bibinfo {year} {1968})}\BibitemShut {NoStop}%
\bibitem [{\citenamefont {Tichovolsky}\ and\ \citenamefont
  {Mavroides}(1969)}]{Tichovolsky1969}%
  \BibitemOpen
  \bibfield  {author} {\bibinfo {author} {\bibfnamefont {E.}~\bibnamefont
  {Tichovolsky}}\ and\ \bibinfo {author} {\bibfnamefont {J.}~\bibnamefont
  {Mavroides}},\ }\bibfield  {title} {\bibinfo {title} {Magnetoreflection
  studies on the band structure of bismuth-antimony alloys},\ }\href
  {https://doi.org/http://dx.doi.org/10.1016/0038-1098(69)90544-4} {\bibfield
  {journal} {\bibinfo  {journal} {Solid State Communications}\ }\textbf
  {\bibinfo {volume} {7}},\ \bibinfo {pages} {927 } (\bibinfo {year}
  {1969})}\BibitemShut {NoStop}%
\bibitem [{\citenamefont {Brandt}\ \emph {et~al.}(1972)\citenamefont {Brandt},
  \citenamefont {Ponomarev},\ and\ \citenamefont {Chudinov}}]{Brandt1972}%
  \BibitemOpen
  \bibfield  {author} {\bibinfo {author} {\bibfnamefont {N.~B.}\ \bibnamefont
  {Brandt}}, \bibinfo {author} {\bibfnamefont {Y.~G.}\ \bibnamefont
  {Ponomarev}},\ and\ \bibinfo {author} {\bibfnamefont {S.~M.}\ \bibnamefont
  {Chudinov}},\ }\bibfield  {title} {\bibinfo {title} {Investigation of the
  gapless state in bismuth-antimony alloys},\ }\href
  {https://doi.org/10.1007/BF00653870} {\bibfield  {journal} {\bibinfo
  {journal} {Journal of Low Temperature Physics}\ }\textbf {\bibinfo {volume}
  {8}},\ \bibinfo {pages} {369} (\bibinfo {year} {1972})}\BibitemShut {NoStop}%
\bibitem [{\citenamefont {Oelgart}\ and\ \citenamefont
  {Herrmann}(1976)}]{Oelgart1976}%
  \BibitemOpen
  \bibfield  {author} {\bibinfo {author} {\bibfnamefont {G.}~\bibnamefont
  {Oelgart}}\ and\ \bibinfo {author} {\bibfnamefont {R.}~\bibnamefont
  {Herrmann}},\ }\bibfield  {title} {\bibinfo {title} {Cyclotron masses in
  semiconducting bi$_{1-x}$sb$_x$ alloys},\ }\href
  {https://doi.org/10.1002/pssb.2220750119} {\bibfield  {journal} {\bibinfo
  {journal} {physica status solidi (b)}\ }\textbf {\bibinfo {volume} {75}},\
  \bibinfo {pages} {189} (\bibinfo {year} {1976})}\BibitemShut {NoStop}%
\bibitem [{\citenamefont {Lenoir}\ \emph {et~al.}(1996)\citenamefont {Lenoir},
  \citenamefont {Cassart}, \citenamefont {Michenaud}, \citenamefont
  {Scherrer},\ and\ \citenamefont {Scherrer}}]{Lenoir1996}%
  \BibitemOpen
  \bibfield  {author} {\bibinfo {author} {\bibfnamefont {B.}~\bibnamefont
  {Lenoir}}, \bibinfo {author} {\bibfnamefont {M.}~\bibnamefont {Cassart}},
  \bibinfo {author} {\bibfnamefont {J.-P.}\ \bibnamefont {Michenaud}}, \bibinfo
  {author} {\bibfnamefont {H.}~\bibnamefont {Scherrer}},\ and\ \bibinfo
  {author} {\bibfnamefont {S.}~\bibnamefont {Scherrer}},\ }\href@noop {}
  {\bibfield  {journal} {\bibinfo  {journal} {J. Phys. Chem. Solids}\ }\textbf
  {\bibinfo {volume} {57}},\ \bibinfo {pages} {89} (\bibinfo {year}
  {1996})}\BibitemShut {NoStop}%
\bibitem [{\citenamefont {Fu}\ and\ \citenamefont {Kane}(2007)}]{Fu2007}%
  \BibitemOpen
  \bibfield  {author} {\bibinfo {author} {\bibfnamefont {L.}~\bibnamefont
  {Fu}}\ and\ \bibinfo {author} {\bibfnamefont {C.~L.}\ \bibnamefont {Kane}},\
  }\bibfield  {title} {\bibinfo {title} {Topological insulators with inversion
  symmetry},\ }\href@noop {} {\bibfield  {journal} {\bibinfo  {journal} {Phys.
  Rev. B}\ }\textbf {\bibinfo {volume} {76}},\ \bibinfo {pages} {045302}
  (\bibinfo {year} {2007})}\BibitemShut {NoStop}%
\bibitem [{\citenamefont {Ast}\ and\ \citenamefont {H\"ochst}(2003)}]{Ast2003}%
  \BibitemOpen
  \bibfield  {author} {\bibinfo {author} {\bibfnamefont {C.~R.}\ \bibnamefont
  {Ast}}\ and\ \bibinfo {author} {\bibfnamefont {H.}~\bibnamefont {H\"ochst}},\
  }\bibfield  {title} {\bibinfo {title} {Electronic structure of a bismuth
  bilayer},\ }\href {https://doi.org/10.1103/PhysRevB.67.113102} {\bibfield
  {journal} {\bibinfo  {journal} {Phys. Rev. B}\ }\textbf {\bibinfo {volume}
  {67}},\ \bibinfo {pages} {113102} (\bibinfo {year} {2003})}\BibitemShut
  {NoStop}%
\bibitem [{\citenamefont {Koroteev}\ \emph {et~al.}(2004)\citenamefont
  {Koroteev}, \citenamefont {Bihlmayer}, \citenamefont {Gayone}, \citenamefont
  {Chulkov}, \citenamefont {Bl\"ugel}, \citenamefont {Echenique},\ and\
  \citenamefont {Hofmann}}]{Koroteev2004}%
  \BibitemOpen
  \bibfield  {author} {\bibinfo {author} {\bibfnamefont {Y.~M.}\ \bibnamefont
  {Koroteev}}, \bibinfo {author} {\bibfnamefont {G.}~\bibnamefont {Bihlmayer}},
  \bibinfo {author} {\bibfnamefont {J.~E.}\ \bibnamefont {Gayone}}, \bibinfo
  {author} {\bibfnamefont {E.~V.}\ \bibnamefont {Chulkov}}, \bibinfo {author}
  {\bibfnamefont {S.}~\bibnamefont {Bl\"ugel}}, \bibinfo {author}
  {\bibfnamefont {P.~M.}\ \bibnamefont {Echenique}},\ and\ \bibinfo {author}
  {\bibfnamefont {P.}~\bibnamefont {Hofmann}},\ }\bibfield  {title} {\bibinfo
  {title} {Strong spin-orbit splitting on bi surfaces},\ }\href
  {https://doi.org/10.1103/PhysRevLett.93.046403} {\bibfield  {journal}
  {\bibinfo  {journal} {Phys. Rev. Lett.}\ }\textbf {\bibinfo {volume} {93}},\
  \bibinfo {pages} {046403} (\bibinfo {year} {2004})}\BibitemShut {NoStop}%
\bibitem [{\citenamefont {Hirahara}\ \emph {et~al.}(2006)\citenamefont
  {Hirahara}, \citenamefont {Nagao}, \citenamefont {Matsuda}, \citenamefont
  {Bihlmayer}, \citenamefont {Chulkov}, \citenamefont {Koroteev}, \citenamefont
  {Echenique}, \citenamefont {Saito},\ and\ \citenamefont
  {Hasegawa}}]{Hirahara2006}%
  \BibitemOpen
  \bibfield  {author} {\bibinfo {author} {\bibfnamefont {T.}~\bibnamefont
  {Hirahara}}, \bibinfo {author} {\bibfnamefont {T.}~\bibnamefont {Nagao}},
  \bibinfo {author} {\bibfnamefont {I.}~\bibnamefont {Matsuda}}, \bibinfo
  {author} {\bibfnamefont {G.}~\bibnamefont {Bihlmayer}}, \bibinfo {author}
  {\bibfnamefont {E.~V.}\ \bibnamefont {Chulkov}}, \bibinfo {author}
  {\bibfnamefont {Y.~M.}\ \bibnamefont {Koroteev}}, \bibinfo {author}
  {\bibfnamefont {P.~M.}\ \bibnamefont {Echenique}}, \bibinfo {author}
  {\bibfnamefont {M.}~\bibnamefont {Saito}},\ and\ \bibinfo {author}
  {\bibfnamefont {S.}~\bibnamefont {Hasegawa}},\ }\bibfield  {title} {\bibinfo
  {title} {Role of spin-orbit coupling and hybridization effects in the
  electronic structure of ultrathin bi films},\ }\href
  {https://doi.org/10.1103/PhysRevLett.97.146803} {\bibfield  {journal}
  {\bibinfo  {journal} {Phys. Rev. Lett.}\ }\textbf {\bibinfo {volume} {97}},\
  \bibinfo {pages} {146803} (\bibinfo {year} {2006})}\BibitemShut {NoStop}%
\bibitem [{\citenamefont {Guo}\ \emph {et~al.}(2011)\citenamefont {Guo},
  \citenamefont {Sugawara}, \citenamefont {Takayama}, \citenamefont {Souma},
  \citenamefont {Sato}, \citenamefont {Satoh}, \citenamefont {Ohnishi},
  \citenamefont {Kitaura}, \citenamefont {Sasaki}, \citenamefont {Xue},\ and\
  \citenamefont {Takahashi}}]{HGuo2011}%
  \BibitemOpen
  \bibfield  {author} {\bibinfo {author} {\bibfnamefont {H.}~\bibnamefont
  {Guo}}, \bibinfo {author} {\bibfnamefont {K.}~\bibnamefont {Sugawara}},
  \bibinfo {author} {\bibfnamefont {A.}~\bibnamefont {Takayama}}, \bibinfo
  {author} {\bibfnamefont {S.}~\bibnamefont {Souma}}, \bibinfo {author}
  {\bibfnamefont {T.}~\bibnamefont {Sato}}, \bibinfo {author} {\bibfnamefont
  {N.}~\bibnamefont {Satoh}}, \bibinfo {author} {\bibfnamefont
  {A.}~\bibnamefont {Ohnishi}}, \bibinfo {author} {\bibfnamefont
  {M.}~\bibnamefont {Kitaura}}, \bibinfo {author} {\bibfnamefont
  {M.}~\bibnamefont {Sasaki}}, \bibinfo {author} {\bibfnamefont {Q.-K.}\
  \bibnamefont {Xue}},\ and\ \bibinfo {author} {\bibfnamefont {T.}~\bibnamefont
  {Takahashi}},\ }\bibfield  {title} {\bibinfo {title} {Evolution of surface
  states in bi${}_{1-x}$sb${}_{x}$ alloys across the topological phase
  transition},\ }\href {https://doi.org/10.1103/PhysRevB.83.201104} {\bibfield
  {journal} {\bibinfo  {journal} {Phys. Rev. B}\ }\textbf {\bibinfo {volume}
  {83}},\ \bibinfo {pages} {201104} (\bibinfo {year} {2011})}\BibitemShut
  {NoStop}%
\bibitem [{\citenamefont {Nakamura}\ \emph {et~al.}(2011)\citenamefont
  {Nakamura}, \citenamefont {Kousa}, \citenamefont {Taskin}, \citenamefont
  {Takeichi}, \citenamefont {Nishide}, \citenamefont {Kakizaki}, \citenamefont
  {D'Angelo}, \citenamefont {Lefevre}, \citenamefont {Bertran}, \citenamefont
  {Taleb-Ibrahimi}, \citenamefont {Komori}, \citenamefont {Kimura},
  \citenamefont {Kondo}, \citenamefont {Ando},\ and\ \citenamefont
  {Matsuda}}]{Nakamura2011}%
  \BibitemOpen
  \bibfield  {author} {\bibinfo {author} {\bibfnamefont {F.}~\bibnamefont
  {Nakamura}}, \bibinfo {author} {\bibfnamefont {Y.}~\bibnamefont {Kousa}},
  \bibinfo {author} {\bibfnamefont {A.~A.}\ \bibnamefont {Taskin}}, \bibinfo
  {author} {\bibfnamefont {Y.}~\bibnamefont {Takeichi}}, \bibinfo {author}
  {\bibfnamefont {A.}~\bibnamefont {Nishide}}, \bibinfo {author} {\bibfnamefont
  {A.}~\bibnamefont {Kakizaki}}, \bibinfo {author} {\bibfnamefont
  {M.}~\bibnamefont {D'Angelo}}, \bibinfo {author} {\bibfnamefont
  {P.}~\bibnamefont {Lefevre}}, \bibinfo {author} {\bibfnamefont
  {F.}~\bibnamefont {Bertran}}, \bibinfo {author} {\bibfnamefont
  {A.}~\bibnamefont {Taleb-Ibrahimi}}, \bibinfo {author} {\bibfnamefont
  {F.}~\bibnamefont {Komori}}, \bibinfo {author} {\bibfnamefont {S.-i.}\
  \bibnamefont {Kimura}}, \bibinfo {author} {\bibfnamefont {H.}~\bibnamefont
  {Kondo}}, \bibinfo {author} {\bibfnamefont {Y.}~\bibnamefont {Ando}},\ and\
  \bibinfo {author} {\bibfnamefont {I.}~\bibnamefont {Matsuda}},\ }\bibfield
  {title} {\bibinfo {title} {Topological transition in bi$_{1-x}$sb$_{x}$
  studied as a function of sb doping},\ }\href
  {https://doi.org/10.1103/PhysRevB.84.235308} {\bibfield  {journal} {\bibinfo
  {journal} {Phys. Rev. B}\ }\textbf {\bibinfo {volume} {84}},\ \bibinfo
  {pages} {235308} (\bibinfo {year} {2011})}\BibitemShut {NoStop}%
\bibitem [{\citenamefont {Ohtsubo}\ \emph {et~al.}(2013)\citenamefont
  {Ohtsubo}, \citenamefont {Perfetti}, \citenamefont {Goerbig}, \citenamefont
  {Fevre}, \citenamefont {Bertran},\ and\ \citenamefont
  {Taleb-Ibrahimi}}]{Ohtsubo2013}%
  \BibitemOpen
  \bibfield  {author} {\bibinfo {author} {\bibfnamefont {Y.}~\bibnamefont
  {Ohtsubo}}, \bibinfo {author} {\bibfnamefont {L.}~\bibnamefont {Perfetti}},
  \bibinfo {author} {\bibfnamefont {M.~O.}\ \bibnamefont {Goerbig}}, \bibinfo
  {author} {\bibfnamefont {P.~L.}\ \bibnamefont {Fevre}}, \bibinfo {author}
  {\bibfnamefont {F.}~\bibnamefont {Bertran}},\ and\ \bibinfo {author}
  {\bibfnamefont {A.}~\bibnamefont {Taleb-Ibrahimi}},\ }\bibfield  {title}
  {\bibinfo {title} {Non-trivial surface-band dispersion on bi(111)},\ }\href
  {http://stacks.iop.org/1367-2630/15/i=3/a=033041} {\bibfield  {journal}
  {\bibinfo  {journal} {New Journal of Physics}\ }\textbf {\bibinfo {volume}
  {15}},\ \bibinfo {pages} {033041} (\bibinfo {year} {2013})}\BibitemShut
  {NoStop}%
\bibitem [{\citenamefont {Benia}\ \emph {et~al.}(2015)\citenamefont {Benia},
  \citenamefont {Stra\ss{}er}, \citenamefont {Kern},\ and\ \citenamefont
  {Ast}}]{Benia2015}%
  \BibitemOpen
  \bibfield  {author} {\bibinfo {author} {\bibfnamefont {H.~M.}\ \bibnamefont
  {Benia}}, \bibinfo {author} {\bibfnamefont {C.}~\bibnamefont {Stra\ss{}er}},
  \bibinfo {author} {\bibfnamefont {K.}~\bibnamefont {Kern}},\ and\ \bibinfo
  {author} {\bibfnamefont {C.~R.}\ \bibnamefont {Ast}},\ }\bibfield  {title}
  {\bibinfo {title} {Surface band structure of
  ${\mathrm{bi}}_{1\ensuremath{-}x}{\mathrm{sb}}_{x}(111)$},\ }\href
  {https://doi.org/10.1103/PhysRevB.91.161406} {\bibfield  {journal} {\bibinfo
  {journal} {Phys. Rev. B}\ }\textbf {\bibinfo {volume} {91}},\ \bibinfo
  {pages} {161406} (\bibinfo {year} {2015})}\BibitemShut {NoStop}%
\bibitem [{\citenamefont {Ito}\ \emph {et~al.}(2016)\citenamefont {Ito},
  \citenamefont {Feng}, \citenamefont {Arita}, \citenamefont {Takayama},
  \citenamefont {Liu}, \citenamefont {Someya}, \citenamefont {Chen},
  \citenamefont {Iimori}, \citenamefont {Namatame}, \citenamefont {Taniguchi},
  \citenamefont {Cheng}, \citenamefont {Tang}, \citenamefont {Komori},
  \citenamefont {Kobayashi}, \citenamefont {Chiang},\ and\ \citenamefont
  {Matsuda}}]{Ito2016}%
  \BibitemOpen
  \bibfield  {author} {\bibinfo {author} {\bibfnamefont {S.}~\bibnamefont
  {Ito}}, \bibinfo {author} {\bibfnamefont {B.}~\bibnamefont {Feng}}, \bibinfo
  {author} {\bibfnamefont {M.}~\bibnamefont {Arita}}, \bibinfo {author}
  {\bibfnamefont {A.}~\bibnamefont {Takayama}}, \bibinfo {author}
  {\bibfnamefont {R.-Y.}\ \bibnamefont {Liu}}, \bibinfo {author} {\bibfnamefont
  {T.}~\bibnamefont {Someya}}, \bibinfo {author} {\bibfnamefont {W.-C.}\
  \bibnamefont {Chen}}, \bibinfo {author} {\bibfnamefont {T.}~\bibnamefont
  {Iimori}}, \bibinfo {author} {\bibfnamefont {H.}~\bibnamefont {Namatame}},
  \bibinfo {author} {\bibfnamefont {M.}~\bibnamefont {Taniguchi}}, \bibinfo
  {author} {\bibfnamefont {C.-M.}\ \bibnamefont {Cheng}}, \bibinfo {author}
  {\bibfnamefont {S.-J.}\ \bibnamefont {Tang}}, \bibinfo {author}
  {\bibfnamefont {F.}~\bibnamefont {Komori}}, \bibinfo {author} {\bibfnamefont
  {K.}~\bibnamefont {Kobayashi}}, \bibinfo {author} {\bibfnamefont {T.-C.}\
  \bibnamefont {Chiang}},\ and\ \bibinfo {author} {\bibfnamefont
  {I.}~\bibnamefont {Matsuda}},\ }\bibfield  {title} {\bibinfo {title} {Proving
  nontrivial topology of pure bismuth by quantum confinement},\ }\href
  {https://doi.org/10.1103/PhysRevLett.117.236402} {\bibfield  {journal}
  {\bibinfo  {journal} {Phys. Rev. Lett.}\ }\textbf {\bibinfo {volume} {117}},\
  \bibinfo {pages} {236402} (\bibinfo {year} {2016})}\BibitemShut {NoStop}%
\bibitem [{\citenamefont {Golin}(1968)}]{Golin1968}%
  \BibitemOpen
  \bibfield  {author} {\bibinfo {author} {\bibfnamefont {S.}~\bibnamefont
  {Golin}},\ }\bibfield  {title} {\bibinfo {title} {Band structure of bismuth:
  Pseudopotential approach},\ }\href {https://doi.org/10.1103/PhysRev.166.643}
  {\bibfield  {journal} {\bibinfo  {journal} {Phys. Rev.}\ }\textbf {\bibinfo
  {volume} {166}},\ \bibinfo {pages} {643} (\bibinfo {year}
  {1968})}\BibitemShut {NoStop}%
\bibitem [{\citenamefont {Aguilera}\ \emph {et~al.}(2015)\citenamefont
  {Aguilera}, \citenamefont {Friedrich},\ and\ \citenamefont
  {Bl\"ugel}}]{Aguilera2015}%
  \BibitemOpen
  \bibfield  {author} {\bibinfo {author} {\bibfnamefont {I.}~\bibnamefont
  {Aguilera}}, \bibinfo {author} {\bibfnamefont {C.}~\bibnamefont
  {Friedrich}},\ and\ \bibinfo {author} {\bibfnamefont {S.}~\bibnamefont
  {Bl\"ugel}},\ }\bibfield  {title} {\bibinfo {title} {Electronic phase
  transitions of bismuth under strain from relativistic self-consistent $gw$
  calculations},\ }\href {https://doi.org/10.1103/PhysRevB.91.125129}
  {\bibfield  {journal} {\bibinfo  {journal} {Phys. Rev. B}\ }\textbf {\bibinfo
  {volume} {91}},\ \bibinfo {pages} {125129} (\bibinfo {year}
  {2015})}\BibitemShut {NoStop}%
\bibitem [{\citenamefont {Teo}\ \emph {et~al.}(2008)\citenamefont {Teo},
  \citenamefont {Fu},\ and\ \citenamefont {Kane}}]{Teo2008}%
  \BibitemOpen
  \bibfield  {author} {\bibinfo {author} {\bibfnamefont {J.~C.~Y.}\
  \bibnamefont {Teo}}, \bibinfo {author} {\bibfnamefont {L.}~\bibnamefont
  {Fu}},\ and\ \bibinfo {author} {\bibfnamefont {C.~L.}\ \bibnamefont {Kane}},\
  }\bibfield  {title} {\bibinfo {title} {Surface states and topological
  invariants in three-dimensional topological insulators: Application to
  bi$_{1-x}$sb$_x$},\ }\href {https://doi.org/10.1103/PhysRevB.78.045426}
  {\bibfield  {journal} {\bibinfo  {journal} {Phys. Rev. B}\ }\textbf {\bibinfo
  {volume} {78}},\ \bibinfo {pages} {045426} (\bibinfo {year}
  {2008})}\BibitemShut {NoStop}%
\bibitem [{\citenamefont {Fuseya}\ \emph
  {et~al.}(2015{\natexlab{b}})\citenamefont {Fuseya}, \citenamefont {Zhu},
  \citenamefont {Fauqu\'e}, \citenamefont {Kang}, \citenamefont {Lenoir},\ and\
  \citenamefont {Behnia}}]{Fuseya2015b}%
  \BibitemOpen
  \bibfield  {author} {\bibinfo {author} {\bibfnamefont {Y.}~\bibnamefont
  {Fuseya}}, \bibinfo {author} {\bibfnamefont {Z.}~\bibnamefont {Zhu}},
  \bibinfo {author} {\bibfnamefont {B.}~\bibnamefont {Fauqu\'e}}, \bibinfo
  {author} {\bibfnamefont {W.}~\bibnamefont {Kang}}, \bibinfo {author}
  {\bibfnamefont {B.}~\bibnamefont {Lenoir}},\ and\ \bibinfo {author}
  {\bibfnamefont {K.}~\bibnamefont {Behnia}},\ }\bibfield  {title} {\bibinfo
  {title} {Origin of the large anisotropic $g$ factor of holes in bismuth},\
  }\href {https://doi.org/10.1103/PhysRevLett.115.216401} {\bibfield  {journal}
  {\bibinfo  {journal} {Phys. Rev. Lett.}\ }\textbf {\bibinfo {volume} {115}},\
  \bibinfo {pages} {216401} (\bibinfo {year} {2015}{\natexlab{b}})}\BibitemShut
  {NoStop}%
\bibitem [{See()}]{SeeSM}%
  \BibitemOpen
  \href@noop {} {}\bibinfo {note} {See Supplemental Material [url] for
  details.}\BibitemShut {Stop}%
\bibitem [{\citenamefont {Saito}\ \emph {et~al.}(2016)\citenamefont {Saito},
  \citenamefont {Sawahata}, \citenamefont {Komine},\ and\ \citenamefont
  {Aono}}]{Saito2016}%
  \BibitemOpen
  \bibfield  {author} {\bibinfo {author} {\bibfnamefont {K.}~\bibnamefont
  {Saito}}, \bibinfo {author} {\bibfnamefont {H.}~\bibnamefont {Sawahata}},
  \bibinfo {author} {\bibfnamefont {T.}~\bibnamefont {Komine}},\ and\ \bibinfo
  {author} {\bibfnamefont {T.}~\bibnamefont {Aono}},\ }\bibfield  {title}
  {\bibinfo {title} {Tight-binding theory of surface spin states on bismuth
  thin films},\ }\href {https://doi.org/10.1103/PhysRevB.93.041301} {\bibfield
  {journal} {\bibinfo  {journal} {Phys. Rev. B}\ }\textbf {\bibinfo {volume}
  {93}},\ \bibinfo {pages} {041301} (\bibinfo {year} {2016})}\BibitemShut
  {NoStop}%
\bibitem [{\citenamefont {Chen}\ \emph {et~al.}(2010)\citenamefont {Chen},
  \citenamefont {Liu}, \citenamefont {Analytis}, \citenamefont {Chu},
  \citenamefont {Zhang}, \citenamefont {Yan}, \citenamefont {Mo}, \citenamefont
  {Moore}, \citenamefont {Lu}, \citenamefont {Fisher}, \citenamefont {Zhang},
  \citenamefont {Hussain},\ and\ \citenamefont {Shen}}]{Chen2010}%
  \BibitemOpen
  \bibfield  {author} {\bibinfo {author} {\bibfnamefont {Y.~L.}\ \bibnamefont
  {Chen}}, \bibinfo {author} {\bibfnamefont {Z.~K.}\ \bibnamefont {Liu}},
  \bibinfo {author} {\bibfnamefont {J.~G.}\ \bibnamefont {Analytis}}, \bibinfo
  {author} {\bibfnamefont {J.-H.}\ \bibnamefont {Chu}}, \bibinfo {author}
  {\bibfnamefont {H.~J.}\ \bibnamefont {Zhang}}, \bibinfo {author}
  {\bibfnamefont {B.~H.}\ \bibnamefont {Yan}}, \bibinfo {author} {\bibfnamefont
  {S.-K.}\ \bibnamefont {Mo}}, \bibinfo {author} {\bibfnamefont {R.~G.}\
  \bibnamefont {Moore}}, \bibinfo {author} {\bibfnamefont {D.~H.}\ \bibnamefont
  {Lu}}, \bibinfo {author} {\bibfnamefont {I.~R.}\ \bibnamefont {Fisher}},
  \bibinfo {author} {\bibfnamefont {S.~C.}\ \bibnamefont {Zhang}}, \bibinfo
  {author} {\bibfnamefont {Z.}~\bibnamefont {Hussain}},\ and\ \bibinfo {author}
  {\bibfnamefont {Z.-X.}\ \bibnamefont {Shen}},\ }\bibfield  {title} {\bibinfo
  {title} {Single dirac cone topological surface state and unusual
  thermoelectric property of compounds from a new topological insulator
  family},\ }\href {https://doi.org/10.1103/PhysRevLett.105.266401} {\bibfield
  {journal} {\bibinfo  {journal} {Phys. Rev. Lett.}\ }\textbf {\bibinfo
  {volume} {105}},\ \bibinfo {pages} {266401} (\bibinfo {year}
  {2010})}\BibitemShut {NoStop}%
\bibitem [{\citenamefont {Eremeev}\ \emph {et~al.}(2011)\citenamefont
  {Eremeev}, \citenamefont {Bihlmayer}, \citenamefont {Vergniory},
  \citenamefont {Koroteev}, \citenamefont {Menshchikova}, \citenamefont {Henk},
  \citenamefont {Ernst},\ and\ \citenamefont {Chulkov}}]{Eremeev2011}%
  \BibitemOpen
  \bibfield  {author} {\bibinfo {author} {\bibfnamefont {S.~V.}\ \bibnamefont
  {Eremeev}}, \bibinfo {author} {\bibfnamefont {G.}~\bibnamefont {Bihlmayer}},
  \bibinfo {author} {\bibfnamefont {M.}~\bibnamefont {Vergniory}}, \bibinfo
  {author} {\bibfnamefont {Y.~M.}\ \bibnamefont {Koroteev}}, \bibinfo {author}
  {\bibfnamefont {T.~V.}\ \bibnamefont {Menshchikova}}, \bibinfo {author}
  {\bibfnamefont {J.}~\bibnamefont {Henk}}, \bibinfo {author} {\bibfnamefont
  {A.}~\bibnamefont {Ernst}},\ and\ \bibinfo {author} {\bibfnamefont {E.~V.}\
  \bibnamefont {Chulkov}},\ }\bibfield  {title} {\bibinfo {title} {Ab initio
  electronic structure of thallium-based topological insulators},\ }\href
  {https://doi.org/10.1103/PhysRevB.83.205129} {\bibfield  {journal} {\bibinfo
  {journal} {Phys. Rev. B}\ }\textbf {\bibinfo {volume} {83}},\ \bibinfo
  {pages} {205129} (\bibinfo {year} {2011})}\BibitemShut {NoStop}%
\bibitem [{\citenamefont {Paraskevopoulos}(1985)}]{Paraskevopoulos1985}%
  \BibitemOpen
  \bibfield  {author} {\bibinfo {author} {\bibfnamefont {K.~M.}\ \bibnamefont
  {Paraskevopoulos}},\ }\bibfield  {title} {\bibinfo {title} {Electron
  effective mass dependence on carrier concentration in tibite2 monocrystals},\
  }\href {https://doi.org/https://doi.org/10.1002/pssb.2221270156} {\bibfield
  {journal} {\bibinfo  {journal} {physica status solidi (b)}\ }\textbf
  {\bibinfo {volume} {127}},\ \bibinfo {pages} {K45} (\bibinfo {year}
  {1985})}\BibitemShut {NoStop}%
\bibitem [{\citenamefont {Michiardi}\ \emph {et~al.}(2014)\citenamefont
  {Michiardi}, \citenamefont {Aguilera}, \citenamefont {Bianchi}, \citenamefont
  {de~Carvalho}, \citenamefont {Ladeira}, \citenamefont {Teixeira},
  \citenamefont {Soares}, \citenamefont {Friedrich}, \citenamefont {Bl\"ugel},\
  and\ \citenamefont {Hofmann}}]{Michiardi2014}%
  \BibitemOpen
  \bibfield  {author} {\bibinfo {author} {\bibfnamefont {M.}~\bibnamefont
  {Michiardi}}, \bibinfo {author} {\bibfnamefont {I.}~\bibnamefont {Aguilera}},
  \bibinfo {author} {\bibfnamefont {M.}~\bibnamefont {Bianchi}}, \bibinfo
  {author} {\bibfnamefont {V.~E.}\ \bibnamefont {de~Carvalho}}, \bibinfo
  {author} {\bibfnamefont {L.~O.}\ \bibnamefont {Ladeira}}, \bibinfo {author}
  {\bibfnamefont {N.~G.}\ \bibnamefont {Teixeira}}, \bibinfo {author}
  {\bibfnamefont {E.~A.}\ \bibnamefont {Soares}}, \bibinfo {author}
  {\bibfnamefont {C.}~\bibnamefont {Friedrich}}, \bibinfo {author}
  {\bibfnamefont {S.}~\bibnamefont {Bl\"ugel}},\ and\ \bibinfo {author}
  {\bibfnamefont {P.}~\bibnamefont {Hofmann}},\ }\bibfield  {title} {\bibinfo
  {title} {Bulk band structure of ${\mathrm{bi}}_{2}{\mathrm{te}}_{3}$},\
  }\href {https://doi.org/10.1103/PhysRevB.90.075105} {\bibfield  {journal}
  {\bibinfo  {journal} {Phys. Rev. B}\ }\textbf {\bibinfo {volume} {90}},\
  \bibinfo {pages} {075105} (\bibinfo {year} {2014})}\BibitemShut {NoStop}%
\end{thebibliography}%

\end{document}